\documentclass[reprint,
superscriptaddress,
nofootinbib,
amsmath,amssymb,
aps,
]{revtex4-1}

\usepackage{tikz}
\usepackage[utf8]{inputenc}
\usepackage{amsmath,amsfonts,amssymb}
\usepackage{graphicx}
\usepackage[toc,page]{appendix}
\usepackage{hyperref}
\usepackage{cleveref}

\usepackage[loose]{units}
\usepackage{booktabs}
\usepackage{float}
\usepackage{mathtools}
\usepackage{multirow}

\definecolor{pyblue}{RGB}{31, 119, 180}

\hypersetup{
    colorlinks=true,
    linkcolor={pyblue},
    citecolor={pyblue},
    urlcolor={pyblue}
}

\makeatletter
\gdef\@fpheader{}
\makeatother

\usepackage{graphicx}
\usepackage{dcolumn}
\usepackage{bm}

\def\MPl{M_{\rm Pl}}
\def\tb{\tilde{b}}
\def\barV{\bar{V}}
\def\F{{\cal F}}
\newcommand{\pbh}{\text{PBH}}
\newcommand{\eq}{\text{eq}}

\def\MPl{M_{\rm Pl}}
\def\tb{\tilde{b}}

\begin{document}
	
	\preprint{}
	
	\title{Inflationary Butterfly Effect:\\ Non-Perturbative Dynamics From Small-Scale Features}
	
	\author{Angelo Caravano}
	\affiliation{Institut d'Astrophysique de Paris, UMR 7095 du CNRS et de Sorbonne Universit\'e,\\ 98 bis bd Arago, 75014 Paris, France}

	\author{Keisuke Inomata}
	
	\affiliation{William H. Miller III Department of Physics and Astronomy,
Johns Hopkins University, 3400 N. Charles Street, Baltimore, Maryland, 21218, USA}
	\author{S\'ebastien Renaux-Petel}
	\affiliation{Institut d'Astrophysique de Paris, UMR 7095 du CNRS et de Sorbonne Universit\'e,\\ 98 bis bd Arago, 75014 Paris, France}

	\begin{abstract}
For the first time, we investigate the non-perturbative dynamics of single field inflation with a departure from slow-roll. Using simulations, we find that oscillatory features in the potential can drastically alter the course of inflation, with major phenomenological implications. In certain cases, the entire Universe gets trapped in a forever inflating de Sitter state. In others, only some regions get stuck in a false vacuum, offering an alternative channel for primordial black hole formation. Analogous to the flap of a butterfly, these results show that small-scale phenomena can have profound consequences on the evolution of the entire Universe. 
More generally, our work shows the power of simulations in the exploration of the small-scale physics of inflation, particularly in the regime relevant for gravitational-wave astronomy.

	\end{abstract}
	
	\maketitle

\section{Introduction}

	Small-scale phenomena can profoundly influence vastly larger natural systems, as famously illustrated by the butterfly effect on the Earth’s atmosphere~\cite{lorenz1972predictability}. This concept emerged from the first computer simulations of convective processes, which laid the foundation for chaos theory~\cite{lorenz1963deterministic}. The interplay between small and large scales can also be important in primordial cosmology.  An example is Linde's chaotic model of inflation~\cite{Linde:1983gd,Linde:1986a}, in which a tiny Planck-scale region of space-time ends up becoming the entire observable Universe.
	In this paper, we use simulations to explore small-scale phenomena occurring during the inflationary epoch. We demonstrate that, similarly to the flap of a butterfly, these phenomena can induce critical nonlinear consequences on the history and properties of the Universe.

	Inflation stands as the most natural mechanism for explaining the origin of structures in the Universe \cite{PhysRevD.23.347,Sato:1980yn,Linde:1981mu,PhysRevLett.48.1220,STAROBINSKY198099,Starobinsky:1979ty,Mukhanov:1981xt,HAWKING1982295,PhysRevLett.49.1110,STAROBINSKY1982175,Abbott:1984fp}. Large-scale observations, such as those of the cosmic microwave background (CMB), tightly constrain the potential of the ``inflaton", the scalar field driving inflation \cite{Planck:2018jri,Planck:2019kim,BICEP:2021xfz}.
	By contrast, density fluctuations on smaller scales, forged in the ``dark era'' of inflation, remain unknown. Theoretically, four decades of work have confirmed that the sensitivity of inflation to Planck-scale physics makes it challenging to realize a long period of slow-roll inflation \cite{Baumann:2014nda}. 
	Hence, inflationary scenarios deviating on small scales from the known large-scale slow-roll dynamics deserve serious scrutiny. 
	Fortunately, the advent of gravitational-wave (GW) astronomy now provides us with a formidable opportunity
	to grasp that unknown cosmological epoch (see \cite{Achucarro:2022qrl,LISACosmologyWorkingGroup:2022jok,LISACosmologyWorkingGroup:2023njw} for recent reviews).

 Deviation from slow-roll generically leads to enhanced density fluctuations. These can seed primordial black holes (PBHs)~\cite{Zeldovich:1967lct,Hawking:1971ei,Carr:1974nx,Carr:1975qj,Chapline:1975ojl}, acting as dark matter, and leave unique imprints in
	the gravitational wave background (GWB) (see the above reviews~\cite{Achucarro:2022qrl,LISACosmologyWorkingGroup:2022jok,LISACosmologyWorkingGroup:2023njw} and references therein). However, these large fluctuations challenge our perturbative understanding of inflation. Actually, many models generate a GWB loud enough to be detectable by future observatories, precisely when perturbative control becomes under threat, see e.g. \cite{Fumagalli:2020nvq,Inomata:2021zel,Inomata:2021tpx,Fumagalli:2021mpc,Inomata:2022yte,Unal:2023srk,Iacconi:2023slv}. 
	Hence, it is crucial that theorists develop non-perturbative methods, to understand the small-scale physics of inflation and to make firm predictions to test against observations.

	In this work, 
	we present the first non-perturbative study of a single-field model of inflation with enhanced density fluctuations.
	We consider one of the simplest and most theoretically motivated class of models,
	single-field inflation with oscillatory features in its potential, that we study using 
	lattice field theory simulations.
	This approach has been extensively used to understand preheating after inflation (see, e.g.,~\cite{Khlebnikov_1996,Prokopec_1997,latticeeasy,Garcia-Bellido:2007nns,Frolov_2008,hlattice,Sainio_2012,Child_2013,Easther_2010,Lozanov_2020,figueroa2021cosmolattice}), and it has more recently been extended to inflation \cite{Caravano:2021pgc,Caravano:2021bfn,Caravano:2022epk,Caravano:2022yyv,Krajewski:2022uuh,Figueroa:2023oxc} (see also, e.g.~, \cite{East:2015ggf,Braden:2016tjn,Clough:2016ymm,Clough:2017efm} for general relativistic treatments).

	We demonstrate that tiny small-scale quantum fluctuations, amplified by the oscillatory feature, can drastically affect the fate of the entire Universe, sometimes preventing inflation from ending.  
	This ``inflationary butterfly effect" reveals the need for a non-perturbative treatment in the regime relevant for observations. 
	This marks the beginning of a new era in the exploration of the Universe's earliest epoch, reminiscent in some ways of the pioneering simulations of chaotic phenomena in \cite{lorenz1963deterministic}.

	\section{Breakdown of perturbation theory}
	\label{sec:model}

	\begin{figure*}
		\centering
		
		\begin{tikzpicture}
			\node (img) {\includegraphics[width=17.8cm]{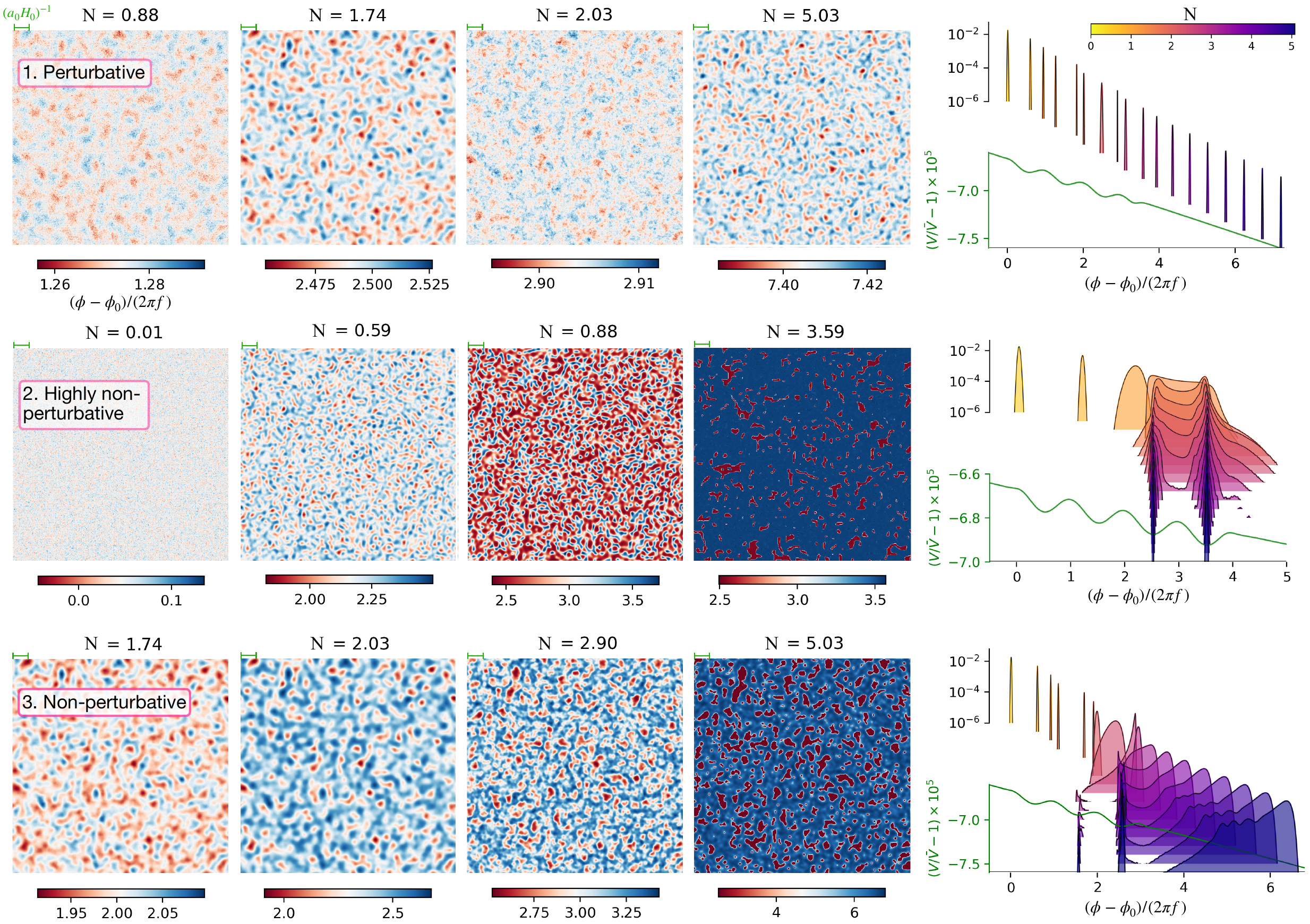}};

		\end{tikzpicture}

		\caption{(Left) 2D slices of the simulations at different times $N=\log(a/a_0)$. (Right) One-point PDF of the inflaton field value at different times, as given by the colorbar in the top-right corner. 
			The PDFs range from $N=0.1$ to $N=5.03$ and are equally spaced by $\Delta N=0.29$.
			They are shown together with the potential (green).  Animations of the snapshots and corresponding PDFs can be found at this \href{https://github.com/caravangelo/Inflationary-Butterfly.git}{link}. }

		\label{fig:snap}
	\end{figure*}
{Inflationary potentials with oscillatory features are a well-known class of models, originally introduced in the context of primordial non-Gaussianity~\cite{Chen:2008wn}, and which can be realized with axion monodromy in string theory~\cite{Silverstein:2008sg,McAllister:2008hb,Flauger:2009ab}.}
We consider a localized oscillation:
	\begin{equation}
		V(\phi)=
		V_{\rm sr}(\phi) 
		+ 
		\Lambda^{4} {\cal W}(\phi) \Bigl[  \cos\left(\frac{\phi-\phi_0}{f}\right) -1 \Bigr],
	\end{equation}
	where $\phi$ is the inflaton, $V_{\rm sr}$ is a slow-roll potential and $\cal{W}$ is a smooth top-hat function that localizes the oscillatory feature between $\phi_0$ and $\phi_0+\Delta\phi$,
	$\mathcal{W(\phi)}=\frac{1}{4}\left(1+\tanh \left(\frac{\phi-\phi_0}{f}\right)\right) \left(1+\tanh\left(\frac{\phi_0-\phi+\Delta\phi}{f}\right)\right)$.
	We set $V_{\rm sr}=\barV \left(1-\frac{1-n_s}{2}\frac{\phi^2}{2 \MPl^2}\right)$, with $\barV$ and $n_s$ chosen to fit the 
	CMB-scale curvature power spectrum away from the feature.

	The amplitude of the oscillation is parametrized by the dimensionless quantity $\tb = 2\Lambda^4/\dot\phi_0^2$ \cite{Creminelli:2024cge}. 
	Here $\dot\phi_0^2=2 \epsilon_0 H_0^2 \MPl^2$, with $\epsilon_0= \frac{\phi_0^2}{2}\left(\frac{1-n_s}{2}\right)^2$ the slow-roll parameter when the inflaton reaches the feature, and $\phi_0$ is a free parameter, determining which scales are affected by the feature. Throughout this study, we set $\epsilon_0=10^{-6}$ as a fiducial value.  The oscillation frequency is $\omega=\dot \phi_0/f \equiv \alpha H_0$, where $H_0$ is the Hubble parameter at $\phi_0$, and $\alpha$ quantifies the number of oscillations per Hubble time.
	
	In the regime of interest with $\alpha \gg 1$, the oscillations resonate
	with the quantum oscillations of fluctuations inside the Hubble radius. This mechanism
	has been extensively studied in the regime of small-amplitude oscillation with $\tilde b\ll 1$, leading to so-called resonant features in the statistics of primordial fluctuations \cite{Chen:2008wn,Flauger:2009ab, Flauger:2010ja,Behbahani:2011it,DuasoPueyo:2023viy,Creminelli:2024cge}.
	In this paper, we consider situations with $\tb \sim 1$, in which the potential is not monotonous. This regime of parameter space has only been studied recently in \cite{Inomata:2022yte}, for the three setups in Table~\ref{tab:parameters}.
	
\begin{table}
	\begin{tabular}{ |c||c|c|c|  }
		
		\hline
		Setup  & $\tb\equiv\frac{2\Lambda^4}{\dot\phi_0^2}$  &$\alpha\equiv\frac{\dot\phi_0}{f H_0}$ &$\frac{\Delta\phi}{\alpha f}$ \\
		\hline
		1   & 1.14    &10&   2.13\\
		2&   1.218  & 25   &1\\
		3 &1.32 & 10&  2\\
		
		\hline
	\end{tabular}
	
	\caption{Three choices of parameters. \label{tab:parameters}}
	
\end{table}

According to standard perturbation theory, in these three cases, the homogeneous part of the inflaton escapes the local minima of the oscillatory potential, i.e.~it goes through the feature and reaches the second slow-roll part of the potential. 
In the first setup, the ``tree-level" power spectrum of the comoving curvature perturbation $\zeta$---as computed from linear perturbation theory---reaches $\mathcal{P}_{\zeta}\simeq 10^{-5}$. In the other two situations it reaches $\mathcal{P}_{\zeta}\simeq 10^{-2}$, the value typically required to efficiently produce PBHs~\cite{Sasaki:2018dmp}. This value is sometimes believed to guarantee a perturbative description. However, this is not the case: as shown in \cite{Inomata:2022yte}, for the last two cases, the backreaction of fluctuations on the inflationary background is significant, and the 1-loop corrected power spectrum significantly differs from the tree-level one, signaling a breakdown of perturbation theory.

\section{Lattice simulations}
The breakdown of perturbation theory implies that one can not separate the inflaton background from its fluctuations. To overcome this, 
we resort to lattice simulations of inflation, using the techniques recently developed in \cite{Caravano:2021pgc,Caravano:2021bfn,Caravano:2022epk,Caravano:2022yyv}, to which we refer for more details. 

In the simulations, we solve the fully nonlinear dynamics of the inflaton field, in a homogeneous Friedmann-Lema\^itre-Robertson-Walker (FLRW) metric whose dynamics is governed by the averaged density and pressure in our box. Working in such a decoupling limit by neglecting metric fluctuations is perfectly legitimate in our situations with $\epsilon\equiv-\dot H/H^2 \ll1$ \cite{Cheung:2007st, Behbahani:2011it,Creminelli:2024cge}. We also stress that, like for preheating \cite{Armendariz-Picon:2019csc}, evolving the system using classical equations of motion is adequate in our regime of large occupation number:
the large amount of particle production, which causes the breakdown of perturbation theory, also guarantees their validity to capture physical effects driven by classical nonlinearities.

The simulations rely on the discretization of space on a grid of $N^3_{\rm pts}$ values, which we set to $N^3_{\rm pts}=512^3$ (unless otherwise specified). The comoving length of the box is set to $L = 12/ (a_0H_0) \equiv 12/k_0$, so that modes captured in the simulations have wavenumbers $0.52 \leq k/k_0 \leq 147$ \cite{Caravano:2021pgc}. We stop the simulations at $N\equiv\log(a/a_0)\simeq 5$ $e$-folds.

\section{Results}
\label{sec:results}

We now present simulation results for the three setups introduced in Table~\ref{tab:parameters}, which we call respectively perturbative, highly non-perturbative and non-perturbative. Our findings are summarized in Fig.~\ref{fig:snap}, which displays 2D snapshots of the simulations along with the one-point probability density function (PDF) of the inflaton at different times, overlaid with a plot of the potential. In Fig.~\ref{fig:powerspectrum}, we show the power spectra at the end of the simulations and compare them to the tree-level and one-loop results in the literature. In the following, we give a more extensive description of these results.

\subsection*{1. Perturbative case}

The first case is in agreement with the perturbative expectation. In the top-right corner of Fig.~\ref{fig:snap}, the PDF shows that the inflaton consists of
a main background value with relatively small fluctuations, which move together along the oscillatory potential and escape the feature.

\begin{figure}
	\centering

	\begin{tikzpicture}
		\node (img) {\includegraphics[width=8.5cm]{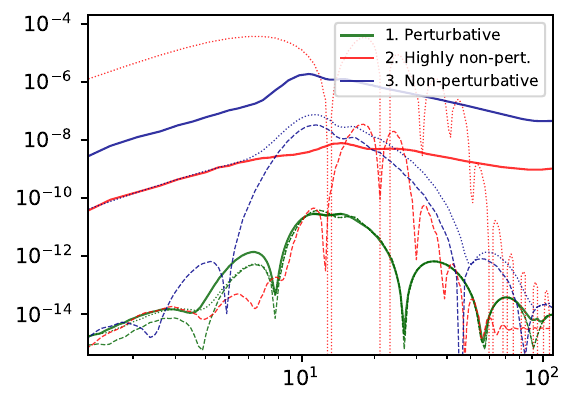}};
		\node at (0.6,3.1){ Power spectrum of the inflaton $\mathcal{P}_\phi$ };
		
		\node [text width=0.01cm,align=center] at (0.5,-3){$k/k_0$};

	\end{tikzpicture}

	\caption{Power spectrum of $\phi$ at $N= 5$. {Solid} lines are the simulation results, while dashed and dotted lines are respectively the tree-level and {1-loop corrected} power spectra, as computed in \cite{Inomata:2022yte} using standard perturbation theory.}
	\label{fig:powerspectrum}
\end{figure}

As shown in Fig.~\ref{fig:powerspectrum}, the final spiky power spectrum agrees remarkably with the perturbative result, including the 1-loop correction in the infrared tail $k <4 k_0$. The only differences are the second peak around $k\simeq 6 k_0$, which may be due to the phenomenon of infrared rescattering \cite{Fumagalli:2023loc} beyond 1-loop, and some minor features around the highest peak. 

The typical size of the structures in the inflaton field is determined by the peak in the power spectrum, roughly given by the resonance scale $k\sim\alpha\, k_0$. This is confirmed by the snapshots of Fig.~\ref{fig:snap}, where $(a_0H_0)^{-1}=k_0^{-1}$, the comoving Hubble radius at the initial time of the feature, serves as a reference scale.

\subsection*{2. Highly non-perturbative case}
The second case is characterized by a larger amplitude and frequency of the oscillation. Contrary to the perturbative expectation, the inflaton gets stuck inside the oscillatory part of the potential. As shown in Fig.~\ref{fig:snap}, the inflaton settles within two local minima of the potential, shown in red and blue in the snapshot at $N=3.59$.
The mechanism behind this trapping phenomenon is as follows: the resonance amplifies the amplitude of typical field fluctuations -- roughly the width of the PDF -- to the point that they become of order the period of the oscillation in the potential. 
However, the gradient energy of this highly inhomogeneous field configuration comes from somewhere: the background kinetic energy of the inflaton. As a consequence, the inflaton cannot escape the local minima, whereas the homogeneous inflaton could.

In addition, Fig.~\ref{fig:snap} shows that some points initially escape the oscillatory region, but they are soon dragged back to the local minima as a consequence of the gradient force, a mechanism analogous to the ``pullback effect of gradients" described in \cite{Clough:2016ymm}. 

We conclude that, due to backreaction, the Universe gets stuck in a de Sitter state. In addition to the results shown in Fig.~\ref{fig:snap}, the \href{https://github.com/caravangelo/Inflationary-Butterfly.git}{animations} show that, after the trapping has occurred at $N\simeq 1.5$, regions in the least populated local minimum (red in Fig.~\ref{fig:snap}) are progressively shrinking. However, the finite grid size does not allow to determine their final sizes in a way that is insensitive to the UV resolution. Understanding the long-term behavior of the false-vacuum trapped Universe would require a detailed study of this system using small-scale simulations, which is beyond the scope of this work. Note also that quantum tunneling might be relevant for the long-term dynamics, and that the validity of lattice simulations for studying vacuum decay processes has been debated
in the recent literature \cite{Braden:2018tky,Hertzberg:2019wgx,Hertzberg:2020tqa,Blanco-Pillado:2019xny,Mou:2019gyl,Ai:2019fri,Michel:2019nwa,Huang:2020bzb,Pirvu:2021roq,Braden:2022odm,Batini:2023zpi,Ai:2023yce,Miyachi:2023fss}.

As perturbation theory breaks down, the tree-level power spectrum is not meaningful. Despite this, note that the lattice power spectrum in~Fig.~\ref{fig:powerspectrum} is much broader than the one-loop spectrum, itself broader than the tree-level one, due to mode couplings.

\subsection*{3. Non-perturbative case}

We now consider the last case, in which the oscillation has the same frequency as the perturbative case but a larger amplitude. Here, only some patches of the Universe are trapped in the oscillatory region of the potential. These patches are shown in red in the final simulation snapshot at $N\simeq5$ in the lower part of Fig.~\ref{fig:snap}. Notice that the comoving sizes of the trapped patches---larger than in the highly non-perturbative case because of the smaller value of $\alpha$---are approximately constant in time. The rest of the Universe keeps inflating in a slow-roll manner, as shown by the PDF in the lower-right corner of Fig.~\ref{fig:snap}. Most of the trapped patches are stuck in the last-encountered local minimum, while only a small fraction is stuck in the previous one.

In the slow-roll inflating regions, non-Gaussianity is large and manifests in the PDF as an exponential tail with superimposed oscillations. This striking behavior may have a similar origin as the oscillations on the tail recently identified in a related context in~\cite{Creminelli:2024cge}.

Finally, in Fig.~\ref{fig:powerspectrum}, we show that the power spectrum is less peaked and is much larger than the perturbative expectation, as a consequence of the large difference in field values between the slow-roll regions and the trapped ones.

\section{Trapped regions as black holes}
\label{sec:pbh}

\begin{figure}
	\centering
	\begin{tikzpicture}
		\node (img) at (1.8,0){\includegraphics[scale=1]{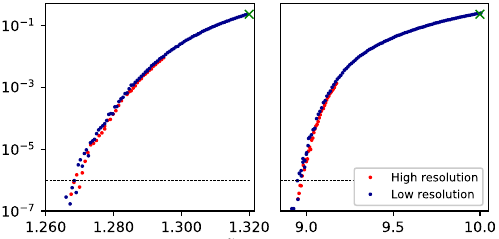}};
		\node at (2.1,2.2){ Trapped volume fraction $\F$ };
		\node at (0.1,-2.2){\footnotesize oscillation amplitude \small $\tb$};
		\node  at (0+4,-2.2){ \footnotesize oscillation frequency \small  $\alpha$};
	\end{tikzpicture}
	\caption{Volume fraction of the trapped regions at the end of the simulation $\F$ as a function of the dimensionless parameters $\tb$ (left) and $\alpha$ (right). The third case of Table~\ref{tab:parameters} is highlighted by a green cross in the top-right corner of each plot. All other points represent simulations with different $\tb$ and $\alpha$, while keeping other parameters fixed. Red and blue correspond to different resolutions, given respectively by $(N_{\rm pts},L\,a_0H_0)=(512,12), $  $(256,10)$. The horizontal black dashed lines show $\mathcal F = 10^{-6}$, which corresponds to the phenomenological upper bound on $\mathcal F$ from PBH overproduction (see the main text).
 }
\label{fig:Rc}
\end{figure}
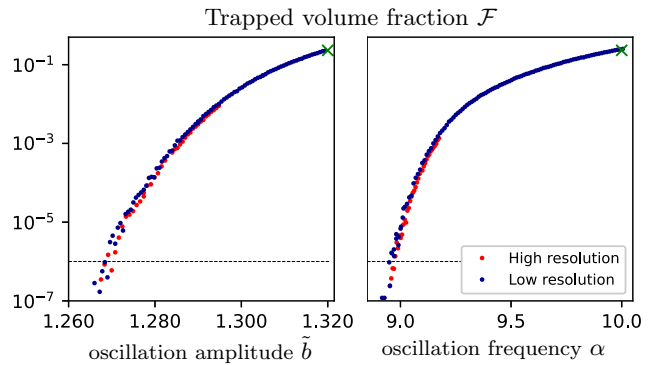

As the first situation considered in Sec.~\ref{sec:results} does not show significant deviation from the perturbative expectation, and the second one exhibits global trapping, ruling out this parameter choice, we now focus on the phenomenology of the third situation, the non-perturbative case.

There, only some patches are stuck in the oscillatory region of the potential. From the point of view of an observer outside these trapped regions, these look as PBHs after the end of inflation. This is similar to what happens in the case of bubble nucleation due to quantum tunneling during inflation~\cite{Deng:2017uwc,Garriga:2015fdk}.
The situation after inflation can be well described in terms of a de Sitter region embedded in a radiation- {or matter-}dominated {(RD or MD)} FLRW spacetime with a lower energy-density. While initially connected through a wormhole, the wormhole later closes, and the de Sitter regions then appear as Schwarzschild black holes for an outside observer \cite{Deng:2017uwc,Garriga:2015fdk,Inomata:2021tpx}.
Note that the trapping phenomenon is different from the one discussed in \cite{Deng:2017uwc,Garriga:2015fdk}: it does not rely on the rare probability to tunnel to other local minima of the potential in the landscape, but instead, the minimum is lying on the classical trajectory.
Such formation of PBHs from trapped regions in single-field models has been already studied in~\cite{Atal:2019cdz,Atal:2019erb,Inomata:2021tpx,Escriva:2023uko,Huang:2023chx}, both analytically and numerically, and is studied here for the first time with 3D lattice simulations.

Let us call $\F=V_{\rm trapped}/V_{\rm total}$ the fraction of the volume in the simulation box in which the inflaton is trapped at the end of the simulation.
Because of causality, $\F$ is preserved during inflation once the trapped regions become larger than the horizon scale $1/(aH)$, which is confirmed by the simulation. 
The mass of PBHs produced from the 
trapped regions is of the same order as the horizon mass at the horizon entry of the bubbles
~\cite{Deng:2017uwc,Garriga:2015fdk}.
Given this, we can roughly regard $\F$ as the mass fraction of PBHs at formation $\beta$~\cite{Inomata:2021tpx}.
The relation between $\beta$ and today's fraction of PBHs over dark matter $f_{\rm PBH}$ depends on the PBH mass and whether they are produced during a MD or RD era.
For example, for solar-mass PBHs produced during RD, $f_{\rm PBH}\sim \mathcal O(1)$ requires $\beta\sim 10^{-9}$~\cite{Sasaki:2018dmp}.\footnote{If we consider supermassive PBHs with, e.g., $M_\pbh \sim 10^{10}\,M_\odot$, $f_\pbh \simeq \mathcal O(1)$ corresponds to $\beta\sim 10^{-4}$, due to the proportionality $f_\pbh \propto M_\pbh^{-1/2}$~\cite{Sasaki:2018dmp}. However, for  $\mathcal O(10^3) < M_\pbh/M_\odot < \mathcal O(10^{11})$,
the passage of nearby PBHs can destroy various astronomical objects, resulting in the bound $f_\pbh \lesssim \mathcal O(10^{-3})$. Hence, even for supermassive PBHs, one should require $\beta \lesssim 10^{-7}$.
}
If we consider PBHs produced during an early MD era, $\beta$ can be increased up to $\sim 10^{-6}$ for $f_{\rm PBH}\sim \mathcal O(1)$, with arbitrary values of the mass (see the Supplemental Material for details).

Given our discussion, and considering PBHs heavier than $\mathcal O(10^{15})\,\text{g}$ that have not evaporated away by now ~\cite{Sasaki:2018dmp}, we require $\F < 10^{-6} $ to avoid PBH overproduction. In the third case considered in \cref{sec:results}, we have $\F \sim 20\%$, which is therefore ruled out. In Fig.~\ref{fig:Rc}, we show the dependencies of the ratio $ \F \simeq \beta$ on the values of the parameters $\tb$ and $\alpha$. We find that the value of $\F$ is extremely sensitive to variations of these parameters. Using these results, we can identify the relevant parameter range to avoid PBH overproduction, corresponding to $\tb < 1.27 $ or $\alpha<9$. In this regime, the simulation still exhibits an exponential tail in the PDF of the inflaton in the slow-roll part of the Universe, which we plan to investigate in future works.

In addition to the trapping phenomenon, PBHs can also form in a conventional manner, as a result of large fluctuations in the slow-roll regions. Estimating this separate contribution to $\beta$ requires computing the comoving curvature perturbation $\zeta$ from the simulation, which has already been done in the context of preheating \cite{Imrith:2018uyk,Imrith:2019njf}. Such a non-perturbative calculation would be a first step towards computing the PBH abundance in non-Gaussian situations, without resorting to simplifying assumptions. We leave this exciting prospect for future work.

\section{Conclusion}

We conducted the first non-perturbative study of inflation with enhanced small-scale perturbations resulting from a departure from slow-roll. Our findings reveal an ``inflationary butterfly effect", where small-scale physical processes drastically alter the evolution of the entire Universe. Notably, these nonlinear effects arise entirely from the quantum nature of inflation, as the dynamics of the inflationary Universe would be unaffected in the absence of small quantum fluctuations.

We focused on the case of an oscillatory feature and demonstrated that the perturbative description completely breaks down while linear perturbation theory predicts $\mathcal P_{\zeta}\sim10^{-2}$. In this case, the inflaton can be trapped in local minima. The trapping can be global, affecting the whole Universe, or local, providing an alternative channel to the formation of PBHs compared to the standard collapse of high-density regions. 
For the latter channel, predicting the PBH abundance from the statistical properties of $\zeta$ is known to be challenging. By contrast, we were able to estimate the abundance of PBHs from trapped patches directly from the simulations.

In the perturbative case, the simulation allowed us to compute inflationary statistics with unprecedented precision. In particular, we compared our fully nonlinear lattice power spectra with perturbative calculations, showing
the first evidence of sizable effects beyond 1-loop, thus motivating analytical developments in this direction.

These results have critical implications for probing the small-scale physics of inflation. In addition to PBHs, 
our findings are crucial for GW astronomy. Indeed, large density
fluctuations 
induce a sizable GWB upon horizon reentry
~\cite{Domenech:2021ztg}.
We expect lattice simulations of inflation to provide invaluable insights into these phenomena, particularly concerning upcoming GW interferometers such as LISA 
\cite{LISA1}
and advanced LIGO \cite{LIGOScientific:2016fpe}, as well as for understanding the possible inflationary origin of the GWB recently detected by pulsar timing array experiments \cite{NANOGrav:2023gor,NANOGrav:2023hvm,EPTA:2023fyk,EPTA:2023xxk,Reardon:2023gzh,Xu:2023wog}.

	\begin{acknowledgments}
We dedicate our work to Ellen Fetter. 
Her numerical calculations in \cite{lorenz1963deterministic} laid the foundation for chaos theory. Her work represents an example of an under-recognized woman in science.

We are particularly grateful to Denis Werth for useful comments and discussions. We also thank Katy Clough, Jaume Garriga, Eiichiro Komatsu, Kaloian D. Lozanov and Arthur Poisson.
A.C. acknowledges funding support from the Initiative Physique des Infinis (IPI), a research training program of the Idex SUPER at Sorbonne Universit\'e.
S.RP is supported by the European Research Council under the European Union’s Horizon 2020 research and innovation programme (grant agreement No
758792, Starting Grant project GEODESI). 
K.I. was supported by JSPS Postdoctoral Fellowships for Research Abroad.
This article is distributed under the Creative Commons Attribution International Licence
(\href{https://creativecommons.org/licenses/by/4.0/}{CC-BY 4.0}).

	\end{acknowledgments}

 	\bibliographystyle{apsrev4-2}
	
	\bibliography{bibliography}

\begin{thebibliography}{99}%
\makeatletter
\providecommand \@ifxundefined [1]{%
 \@ifx{#1\undefined}
}%
\providecommand \@ifnum [1]{%
 \ifnum #1\expandafter \@firstoftwo
 \else \expandafter \@secondoftwo
 \fi
}%
\providecommand \@ifx [1]{%
 \ifx #1\expandafter \@firstoftwo
 \else \expandafter \@secondoftwo
 \fi
}%
\providecommand \natexlab [1]{#1}%
\providecommand \enquote  [1]{``#1''}%
\providecommand \bibnamefont  [1]{#1}%
\providecommand \bibfnamefont [1]{#1}%
\providecommand \citenamefont [1]{#1}%
\providecommand \href@noop [0]{\@secondoftwo}%
\providecommand \href [0]{\begingroup \@sanitize@url \@href}%
\providecommand \@href[1]{\@@startlink{#1}\@@href}%
\providecommand \@@href[1]{\endgroup#1\@@endlink}%
\providecommand \@sanitize@url [0]{\catcode `\\12\catcode `\$12\catcode
  `\&12\catcode `\#12\catcode `\^12\catcode `\_12\catcode `\%12\relax}%
\providecommand \@@startlink[1]{}%
\providecommand \@@endlink[0]{}%
\providecommand \url  [0]{\begingroup\@sanitize@url \@url }%
\providecommand \@url [1]{\endgroup\@href {#1}{\urlprefix }}%
\providecommand \urlprefix  [0]{URL }%
\providecommand \Eprint [0]{\href }%
\providecommand \doibase [0]{https://doi.org/}%
\providecommand \selectlanguage [0]{\@gobble}%
\providecommand \bibinfo  [0]{\@secondoftwo}%
\providecommand \bibfield  [0]{\@secondoftwo}%
\providecommand \translation [1]{[#1]}%
\providecommand \BibitemOpen [0]{}%
\providecommand \bibitemStop [0]{}%
\providecommand \bibitemNoStop [0]{.\EOS\space}%
\providecommand \EOS [0]{\spacefactor3000\relax}%
\providecommand \BibitemShut  [1]{\csname bibitem#1\endcsname}%
\let\auto@bib@innerbib\@empty
\bibitem [{\citenamefont {Lorenz}(1972)}]{lorenz1972predictability}%
  \BibitemOpen
  \bibfield  {author} {\bibinfo {author} {\bibfnamefont {E.~N.}\ \bibnamefont
  {Lorenz}},\ }\href@noop {} {\bibfield  {journal} {\bibinfo  {journal}
  {American Association for the Advancement of Science}\ } (\bibinfo {year}
  {1972})}\BibitemShut {NoStop}%
\bibitem [{\citenamefont {Lorenz}(1963)}]{lorenz1963deterministic}%
  \BibitemOpen
  \bibfield  {author} {\bibinfo {author} {\bibfnamefont {E.~N.}\ \bibnamefont
  {Lorenz}},\ }\href@noop {} {\bibfield  {journal} {\bibinfo  {journal}
  {Journal of atmospheric sciences}\ }\textbf {\bibinfo {volume} {20}},\
  \bibinfo {pages} {130} (\bibinfo {year} {1963})}\BibitemShut {NoStop}%
\bibitem [{\citenamefont {Linde}(1983)}]{Linde:1983gd}%
  \BibitemOpen
  \bibfield  {author} {\bibinfo {author} {\bibfnamefont {A.~D.}\ \bibnamefont
  {Linde}},\ }\href {https://doi.org/10.1016/0370-2693(83)90837-7} {\bibfield
  {journal} {\bibinfo  {journal} {Phys. Lett. B}\ }\textbf {\bibinfo {volume}
  {129}},\ \bibinfo {pages} {177} (\bibinfo {year} {1983})}\BibitemShut
  {NoStop}%
\bibitem [{\citenamefont {Linde}(1986)}]{Linde:1986a}%
  \BibitemOpen
  \bibfield  {author} {\bibinfo {author} {\bibfnamefont {A.~D.}\ \bibnamefont
  {Linde}},\ }\href {https://doi.org/10.1016/0370-2693(86)90611-8} {\bibfield
  {journal} {\bibinfo  {journal} {Phys. Lett. B}\ }\textbf {\bibinfo {volume}
  {175}},\ \bibinfo {pages} {395} (\bibinfo {year} {1986})}\BibitemShut
  {NoStop}%
\bibitem [{\citenamefont {Guth}(1981)}]{PhysRevD.23.347}%
  \BibitemOpen
  \bibfield  {author} {\bibinfo {author} {\bibfnamefont {A.~H.}\ \bibnamefont
  {Guth}},\ }\href {https://doi.org/10.1103/PhysRevD.23.347} {\bibfield
  {journal} {\bibinfo  {journal} {Phys. Rev. D}\ }\textbf {\bibinfo {volume}
  {23}},\ \bibinfo {pages} {347} (\bibinfo {year} {1981})}\BibitemShut
  {NoStop}%
\bibitem [{\citenamefont {Sato}(1981)}]{Sato:1980yn}%
  \BibitemOpen
  \bibfield  {author} {\bibinfo {author} {\bibfnamefont {K.}~\bibnamefont
  {Sato}},\ }\href@noop {} {\bibfield  {journal} {\bibinfo  {journal} {Mon.
  Not. Roy. Astron. Soc.}\ }\textbf {\bibinfo {volume} {195}},\ \bibinfo
  {pages} {467} (\bibinfo {year} {1981})}\BibitemShut {NoStop}%
\bibitem [{\citenamefont {Linde}(1987)}]{Linde:1981mu}%
  \BibitemOpen
  \bibfield  {author} {\bibinfo {author} {\bibfnamefont {A.~D.}\ \bibnamefont
  {Linde}},\ }\href {https://doi.org/10.1016/0370-2693(82)91219-9} {\bibfield
  {journal} {\bibinfo  {journal} {Adv. Ser. Astrophys. Cosmol.}\ }\textbf
  {\bibinfo {volume} {3}},\ \bibinfo {pages} {149} (\bibinfo {year}
  {1987})}\BibitemShut {NoStop}%
\bibitem [{\citenamefont {Albrecht}\ and\ \citenamefont
  {Steinhardt}(1982)}]{PhysRevLett.48.1220}%
  \BibitemOpen
  \bibfield  {author} {\bibinfo {author} {\bibfnamefont {A.}~\bibnamefont
  {Albrecht}}\ and\ \bibinfo {author} {\bibfnamefont {P.~J.}\ \bibnamefont
  {Steinhardt}},\ }\href {https://doi.org/10.1103/PhysRevLett.48.1220}
  {\bibfield  {journal} {\bibinfo  {journal} {Phys. Rev. Lett.}\ }\textbf
  {\bibinfo {volume} {48}},\ \bibinfo {pages} {1220} (\bibinfo {year}
  {1982})}\BibitemShut {NoStop}%
\bibitem [{\citenamefont {Starobinsky}(1980)}]{STAROBINSKY198099}%
  \BibitemOpen
  \bibfield  {author} {\bibinfo {author} {\bibfnamefont {A.}~\bibnamefont
  {Starobinsky}},\ }\href
  {https://doi.org/https://doi.org/10.1016/0370-2693(80)90670-X} {\bibfield
  {journal} {\bibinfo  {journal} {Physics Letters B}\ }\textbf {\bibinfo
  {volume} {91}},\ \bibinfo {pages} {99 } (\bibinfo {year} {1980})}\BibitemShut
  {NoStop}%
\bibitem [{\citenamefont {Starobinsky}(1979)}]{Starobinsky:1979ty}%
  \BibitemOpen
  \bibfield  {author} {\bibinfo {author} {\bibfnamefont {A.~A.}\ \bibnamefont
  {Starobinsky}},\ }\href@noop {} {\bibfield  {journal} {\bibinfo  {journal}
  {JETP Lett.}\ }\textbf {\bibinfo {volume} {30}},\ \bibinfo {pages} {682}
  (\bibinfo {year} {1979})}\BibitemShut {NoStop}%
\bibitem [{\citenamefont {Mukhanov}\ and\ \citenamefont
  {Chibisov}(1981)}]{Mukhanov:1981xt}%
  \BibitemOpen
  \bibfield  {author} {\bibinfo {author} {\bibfnamefont {V.~F.}\ \bibnamefont
  {Mukhanov}}\ and\ \bibinfo {author} {\bibfnamefont {G.~V.}\ \bibnamefont
  {Chibisov}},\ }\href@noop {} {\bibfield  {journal} {\bibinfo  {journal} {JETP
  Lett.}\ }\textbf {\bibinfo {volume} {33}},\ \bibinfo {pages} {532} (\bibinfo
  {year} {1981})}\BibitemShut {NoStop}%
\bibitem [{\citenamefont {Hawking}(1982)}]{HAWKING1982295}%
  \BibitemOpen
  \bibfield  {author} {\bibinfo {author} {\bibfnamefont {S.}~\bibnamefont
  {Hawking}},\ }\href
  {https://doi.org/https://doi.org/10.1016/0370-2693(82)90373-2} {\bibfield
  {journal} {\bibinfo  {journal} {Physics Letters B}\ }\textbf {\bibinfo
  {volume} {115}},\ \bibinfo {pages} {295 } (\bibinfo {year}
  {1982})}\BibitemShut {NoStop}%
\bibitem [{\citenamefont {Guth}\ and\ \citenamefont
  {Pi}(1982)}]{PhysRevLett.49.1110}%
  \BibitemOpen
  \bibfield  {author} {\bibinfo {author} {\bibfnamefont {A.~H.}\ \bibnamefont
  {Guth}}\ and\ \bibinfo {author} {\bibfnamefont {S.-Y.}\ \bibnamefont {Pi}},\
  }\href {https://doi.org/10.1103/PhysRevLett.49.1110} {\bibfield  {journal}
  {\bibinfo  {journal} {Phys. Rev. Lett.}\ }\textbf {\bibinfo {volume} {49}},\
  \bibinfo {pages} {1110} (\bibinfo {year} {1982})}\BibitemShut {NoStop}%
\bibitem [{\citenamefont {Starobinsky}(1982)}]{STAROBINSKY1982175}%
  \BibitemOpen
  \bibfield  {author} {\bibinfo {author} {\bibfnamefont {A.}~\bibnamefont
  {Starobinsky}},\ }\href
  {https://doi.org/https://doi.org/10.1016/0370-2693(82)90541-X} {\bibfield
  {journal} {\bibinfo  {journal} {Physics Letters B}\ }\textbf {\bibinfo
  {volume} {117}},\ \bibinfo {pages} {175 } (\bibinfo {year}
  {1982})}\BibitemShut {NoStop}%
\bibitem [{\citenamefont {Abbott}\ and\ \citenamefont
  {Wise}(1984)}]{Abbott:1984fp}%
  \BibitemOpen
  \bibfield  {author} {\bibinfo {author} {\bibfnamefont {L.}~\bibnamefont
  {Abbott}}\ and\ \bibinfo {author} {\bibfnamefont {M.~B.}\ \bibnamefont
  {Wise}},\ }\href {https://doi.org/10.1016/0550-3213(84)90329-8} {\bibfield
  {journal} {\bibinfo  {journal} {Nucl. Phys. B}\ }\textbf {\bibinfo {volume}
  {244}},\ \bibinfo {pages} {541} (\bibinfo {year} {1984})}\BibitemShut
  {NoStop}%
\bibitem [{\citenamefont {collaboration}(2020)}]{Planck:2018jri}%
  \BibitemOpen
  \bibfield  {author} {\bibinfo {author} {\bibfnamefont {P.}~\bibnamefont
  {collaboration}} (\bibinfo {collaboration} {Planck}),\ }\href
  {https://doi.org/10.1051/0004-6361/201833887} {\bibfield  {journal} {\bibinfo
   {journal} {Astron. Astrophys.}\ }\textbf {\bibinfo {volume} {641}},\
  \bibinfo {pages} {A10} (\bibinfo {year} {2020})},\ \Eprint
  {https://arxiv.org/abs/1807.06211} {arXiv:1807.06211 [astro-ph.CO]}
  \BibitemShut {NoStop}%
\bibitem [{\citenamefont {Akrami}\ \emph {et~al.}(2020)\citenamefont {Akrami}
  \emph {et~al.}}]{Planck:2019kim}%
  \BibitemOpen
  \bibfield  {author} {\bibinfo {author} {\bibfnamefont {Y.}~\bibnamefont
  {Akrami}} \emph {et~al.} (\bibinfo {collaboration} {Planck}),\ }\href
  {https://doi.org/10.1051/0004-6361/201935891} {\bibfield  {journal} {\bibinfo
   {journal} {Astron. Astrophys.}\ }\textbf {\bibinfo {volume} {641}},\
  \bibinfo {pages} {A9} (\bibinfo {year} {2020})},\ \Eprint
  {https://arxiv.org/abs/1905.05697} {arXiv:1905.05697 [astro-ph.CO]}
  \BibitemShut {NoStop}%
\bibitem [{\citenamefont {Ade}\ \emph {et~al.}(2021)\citenamefont {Ade} \emph
  {et~al.}}]{BICEP:2021xfz}%
  \BibitemOpen
  \bibfield  {author} {\bibinfo {author} {\bibfnamefont {P.~A.~R.}\
  \bibnamefont {Ade}} \emph {et~al.} (\bibinfo {collaboration} {BICEP, Keck}),\
  }\href {https://doi.org/10.1103/PhysRevLett.127.151301} {\bibfield  {journal}
  {\bibinfo  {journal} {Phys. Rev. Lett.}\ }\textbf {\bibinfo {volume} {127}},\
  \bibinfo {pages} {151301} (\bibinfo {year} {2021})},\ \Eprint
  {https://arxiv.org/abs/2110.00483} {arXiv:2110.00483 [astro-ph.CO]}
  \BibitemShut {NoStop}%
\bibitem [{\citenamefont {Baumann}\ and\ \citenamefont
  {McAllister}(2015)}]{Baumann:2014nda}%
  \BibitemOpen
  \bibfield  {author} {\bibinfo {author} {\bibfnamefont {D.}~\bibnamefont
  {Baumann}}\ and\ \bibinfo {author} {\bibfnamefont {L.}~\bibnamefont
  {McAllister}},\ }\href {https://doi.org/10.1017/CBO9781316105733} {\emph
  {\bibinfo {title} {{Inflation and String Theory}}}},\ Cambridge Monographs on
  Mathematical Physics\ (\bibinfo  {publisher} {Cambridge University Press},\
  \bibinfo {year} {2015})\ \Eprint {https://arxiv.org/abs/1404.2601}
  {arXiv:1404.2601 [hep-th]} \BibitemShut {NoStop}%
\bibitem [{\citenamefont {Ach\'ucarro}\ \emph {et~al.}(2022)\citenamefont
  {Ach\'ucarro} \emph {et~al.}}]{Achucarro:2022qrl}%
  \BibitemOpen
  \bibfield  {author} {\bibinfo {author} {\bibfnamefont {A.}~\bibnamefont
  {Ach\'ucarro}} \emph {et~al.},\ }\href@noop {} {\  (\bibinfo {year}
  {2022})},\ \Eprint {https://arxiv.org/abs/2203.08128} {arXiv:2203.08128
  [astro-ph.CO]} \BibitemShut {NoStop}%
\bibitem [{\citenamefont {Auclair}\ \emph {et~al.}(2023)\citenamefont {Auclair}
  \emph {et~al.}}]{LISACosmologyWorkingGroup:2022jok}%
  \BibitemOpen
  \bibfield  {author} {\bibinfo {author} {\bibfnamefont {P.}~\bibnamefont
  {Auclair}} \emph {et~al.} (\bibinfo {collaboration} {LISA Cosmology Working
  Group}),\ }\href {https://doi.org/10.1007/s41114-023-00045-2} {\bibfield
  {journal} {\bibinfo  {journal} {Living Rev. Rel.}\ }\textbf {\bibinfo
  {volume} {26}},\ \bibinfo {pages} {5} (\bibinfo {year} {2023})},\ \Eprint
  {https://arxiv.org/abs/2204.05434} {arXiv:2204.05434 [astro-ph.CO]}
  \BibitemShut {NoStop}%
\bibitem [{\citenamefont {Bagui}\ \emph {et~al.}(2023)\citenamefont {Bagui}
  \emph {et~al.}}]{LISACosmologyWorkingGroup:2023njw}%
  \BibitemOpen
  \bibfield  {author} {\bibinfo {author} {\bibfnamefont {E.}~\bibnamefont
  {Bagui}} \emph {et~al.} (\bibinfo {collaboration} {LISA Cosmology Working
  Group}),\ }\href@noop {} {\  (\bibinfo {year} {2023})},\ \Eprint
  {https://arxiv.org/abs/2310.19857} {arXiv:2310.19857 [astro-ph.CO]}
  \BibitemShut {NoStop}%
\bibitem [{\citenamefont {Zel'dovich}\ and\ \citenamefont
  {Novikov}(1967)}]{Zeldovich:1967lct}%
  \BibitemOpen
  \bibfield  {author} {\bibinfo {author} {\bibfnamefont {Y.~B.}\ \bibnamefont
  {Zel'dovich}}\ and\ \bibinfo {author} {\bibfnamefont {I.~D.}\ \bibnamefont
  {Novikov}},\ }\href@noop {} {\bibfield  {journal} {\bibinfo  {journal} {Sov.
  Astron.}\ }\textbf {\bibinfo {volume} {10}},\ \bibinfo {pages} {602}
  (\bibinfo {year} {1967})}\BibitemShut {NoStop}%
\bibitem [{\citenamefont {Hawking}(1971)}]{Hawking:1971ei}%
  \BibitemOpen
  \bibfield  {author} {\bibinfo {author} {\bibfnamefont {S.}~\bibnamefont
  {Hawking}},\ }\href@noop {} {\bibfield  {journal} {\bibinfo  {journal} {Mon.
  Not. Roy. Astron. Soc.}\ }\textbf {\bibinfo {volume} {152}},\ \bibinfo
  {pages} {75} (\bibinfo {year} {1971})}\BibitemShut {NoStop}%
\bibitem [{\citenamefont {Carr}\ and\ \citenamefont
  {Hawking}(1974)}]{Carr:1974nx}%
  \BibitemOpen
  \bibfield  {author} {\bibinfo {author} {\bibfnamefont {B.~J.}\ \bibnamefont
  {Carr}}\ and\ \bibinfo {author} {\bibfnamefont {S.~W.}\ \bibnamefont
  {Hawking}},\ }\href@noop {} {\bibfield  {journal} {\bibinfo  {journal} {Mon.
  Not. Roy. Astron. Soc.}\ }\textbf {\bibinfo {volume} {168}},\ \bibinfo
  {pages} {399} (\bibinfo {year} {1974})}\BibitemShut {NoStop}%
\bibitem [{\citenamefont {Carr}(1975)}]{Carr:1975qj}%
  \BibitemOpen
  \bibfield  {author} {\bibinfo {author} {\bibfnamefont {B.~J.}\ \bibnamefont
  {Carr}},\ }\href {https://doi.org/10.1086/153853} {\bibfield  {journal}
  {\bibinfo  {journal} {Astrophys. J.}\ }\textbf {\bibinfo {volume} {201}},\
  \bibinfo {pages} {1} (\bibinfo {year} {1975})}\BibitemShut {NoStop}%
\bibitem [{\citenamefont {Chapline}(1975)}]{Chapline:1975ojl}%
  \BibitemOpen
  \bibfield  {author} {\bibinfo {author} {\bibfnamefont {G.~F.}\ \bibnamefont
  {Chapline}},\ }\href {https://doi.org/10.1038/253251a0} {\bibfield  {journal}
  {\bibinfo  {journal} {Nature}\ }\textbf {\bibinfo {volume} {253}},\ \bibinfo
  {pages} {251} (\bibinfo {year} {1975})}\BibitemShut {NoStop}%
\bibitem [{\citenamefont {Fumagalli}\ \emph {et~al.}(2021)\citenamefont
  {Fumagalli}, \citenamefont {Renaux-Petel},\ and\ \citenamefont
  {Witkowski}}]{Fumagalli:2020nvq}%
  \BibitemOpen
  \bibfield  {author} {\bibinfo {author} {\bibfnamefont {J.}~\bibnamefont
  {Fumagalli}}, \bibinfo {author} {\bibfnamefont {S.}~\bibnamefont
  {Renaux-Petel}},\ and\ \bibinfo {author} {\bibfnamefont {L.~T.}\ \bibnamefont
  {Witkowski}},\ }\href {https://doi.org/10.1088/1475-7516/2021/08/030}
  {\bibfield  {journal} {\bibinfo  {journal} {JCAP}\ }\textbf {\bibinfo
  {volume} {08}},\ \bibinfo {pages} {030}},\ \Eprint
  {https://arxiv.org/abs/2012.02761} {arXiv:2012.02761 [astro-ph.CO]}
  \BibitemShut {NoStop}%
\bibitem [{\citenamefont {Inomata}(2021)}]{Inomata:2021zel}%
  \BibitemOpen
  \bibfield  {author} {\bibinfo {author} {\bibfnamefont {K.}~\bibnamefont
  {Inomata}},\ }\href {https://doi.org/10.1103/PhysRevD.104.123525} {\bibfield
  {journal} {\bibinfo  {journal} {Phys. Rev. D}\ }\textbf {\bibinfo {volume}
  {104}},\ \bibinfo {pages} {123525} (\bibinfo {year} {2021})},\ \Eprint
  {https://arxiv.org/abs/2109.06192} {arXiv:2109.06192 [astro-ph.CO]}
  \BibitemShut {NoStop}%
\bibitem [{\citenamefont {Inomata}\ \emph {et~al.}(2022)\citenamefont
  {Inomata}, \citenamefont {McDonough},\ and\ \citenamefont
  {Hu}}]{Inomata:2021tpx}%
  \BibitemOpen
  \bibfield  {author} {\bibinfo {author} {\bibfnamefont {K.}~\bibnamefont
  {Inomata}}, \bibinfo {author} {\bibfnamefont {E.}~\bibnamefont {McDonough}},\
  and\ \bibinfo {author} {\bibfnamefont {W.}~\bibnamefont {Hu}},\ }\href
  {https://doi.org/10.1088/1475-7516/2022/02/031} {\bibfield  {journal}
  {\bibinfo  {journal} {JCAP}\ }\textbf {\bibinfo {volume} {02}}\bibfield
  {number} {\bibinfo  {number} { (02)},\ \bibinfo {pages} {031}},\ }\Eprint
  {https://arxiv.org/abs/2110.14641} {arXiv:2110.14641 [astro-ph.CO]}
  \BibitemShut {NoStop}%
\bibitem [{\citenamefont {Fumagalli}\ \emph {et~al.}(2022)\citenamefont
  {Fumagalli}, \citenamefont {Palma}, \citenamefont {Renaux-Petel},
  \citenamefont {Sypsas}, \citenamefont {Witkowski},\ and\ \citenamefont
  {Zenteno}}]{Fumagalli:2021mpc}%
  \BibitemOpen
  \bibfield  {author} {\bibinfo {author} {\bibfnamefont {J.}~\bibnamefont
  {Fumagalli}}, \bibinfo {author} {\bibfnamefont {G.~A.}\ \bibnamefont
  {Palma}}, \bibinfo {author} {\bibfnamefont {S.}~\bibnamefont {Renaux-Petel}},
  \bibinfo {author} {\bibfnamefont {S.}~\bibnamefont {Sypsas}}, \bibinfo
  {author} {\bibfnamefont {L.~T.}\ \bibnamefont {Witkowski}},\ and\ \bibinfo
  {author} {\bibfnamefont {C.}~\bibnamefont {Zenteno}},\ }\href
  {https://doi.org/10.1007/JHEP03(2022)196} {\bibfield  {journal} {\bibinfo
  {journal} {JHEP}\ }\textbf {\bibinfo {volume} {03}},\ \bibinfo {pages}
  {196}},\ \Eprint {https://arxiv.org/abs/2111.14664} {arXiv:2111.14664
  [astro-ph.CO]} \BibitemShut {NoStop}%
\bibitem [{\citenamefont {Inomata}\ \emph {et~al.}(2023)\citenamefont
  {Inomata}, \citenamefont {Braglia}, \citenamefont {Chen},\ and\ \citenamefont
  {Renaux-Petel}}]{Inomata:2022yte}%
  \BibitemOpen
  \bibfield  {author} {\bibinfo {author} {\bibfnamefont {K.}~\bibnamefont
  {Inomata}}, \bibinfo {author} {\bibfnamefont {M.}~\bibnamefont {Braglia}},
  \bibinfo {author} {\bibfnamefont {X.}~\bibnamefont {Chen}},\ and\ \bibinfo
  {author} {\bibfnamefont {S.}~\bibnamefont {Renaux-Petel}},\ }\href
  {https://doi.org/10.1088/1475-7516/2023/04/011} {\bibfield  {journal}
  {\bibinfo  {journal} {JCAP}\ }\textbf {\bibinfo {volume} {04}},\ \bibinfo
  {pages} {011}},\ \bibinfo {note} {[Erratum: JCAP 09, E01 (2023)]},\ \Eprint
  {https://arxiv.org/abs/2211.02586} {arXiv:2211.02586 [astro-ph.CO]}
  \BibitemShut {NoStop}%
\bibitem [{\citenamefont {Unal}\ \emph {et~al.}(2023)\citenamefont {Unal},
  \citenamefont {Papageorgiou},\ and\ \citenamefont {Obata}}]{Unal:2023srk}%
  \BibitemOpen
  \bibfield  {author} {\bibinfo {author} {\bibfnamefont {C.}~\bibnamefont
  {Unal}}, \bibinfo {author} {\bibfnamefont {A.}~\bibnamefont {Papageorgiou}},\
  and\ \bibinfo {author} {\bibfnamefont {I.}~\bibnamefont {Obata}},\
  }\href@noop {} {\  (\bibinfo {year} {2023})},\ \Eprint
  {https://arxiv.org/abs/2307.02322} {arXiv:2307.02322 [astro-ph.CO]}
  \BibitemShut {NoStop}%
\bibitem [{\citenamefont {Iacconi}\ and\ \citenamefont
  {Mulryne}(2023)}]{Iacconi:2023slv}%
  \BibitemOpen
  \bibfield  {author} {\bibinfo {author} {\bibfnamefont {L.}~\bibnamefont
  {Iacconi}}\ and\ \bibinfo {author} {\bibfnamefont {D.~J.}\ \bibnamefont
  {Mulryne}},\ }\href {https://doi.org/10.1088/1475-7516/2023/09/033}
  {\bibfield  {journal} {\bibinfo  {journal} {JCAP}\ }\textbf {\bibinfo
  {volume} {09}},\ \bibinfo {pages} {033}},\ \Eprint
  {https://arxiv.org/abs/2304.14260} {arXiv:2304.14260 [astro-ph.CO]}
  \BibitemShut {NoStop}%
\bibitem [{\citenamefont {Khlebnikov}\ and\ \citenamefont
  {Tkachev}(1996)}]{Khlebnikov_1996}%
  \BibitemOpen
  \bibfield  {author} {\bibinfo {author} {\bibfnamefont {S.~Y.}\ \bibnamefont
  {Khlebnikov}}\ and\ \bibinfo {author} {\bibfnamefont {I.~I.}\ \bibnamefont
  {Tkachev}},\ }\href {https://doi.org/10.1103/physrevlett.77.219} {\bibfield
  {journal} {\bibinfo  {journal} {Physical Review Letters}\ }\textbf {\bibinfo
  {volume} {77}},\ \bibinfo {pages} {219?222} (\bibinfo {year}
  {1996})}\BibitemShut {NoStop}%
\bibitem [{\citenamefont {Prokopec}\ and\ \citenamefont
  {Roos}(1997)}]{Prokopec_1997}%
  \BibitemOpen
  \bibfield  {author} {\bibinfo {author} {\bibfnamefont {T.}~\bibnamefont
  {Prokopec}}\ and\ \bibinfo {author} {\bibfnamefont {T.~G.}\ \bibnamefont
  {Roos}},\ }\href {https://doi.org/10.1103/physrevd.55.3768} {\bibfield
  {journal} {\bibinfo  {journal} {Physical Review D}\ }\textbf {\bibinfo
  {volume} {55}},\ \bibinfo {pages} {3768?3775} (\bibinfo {year}
  {1997})}\BibitemShut {NoStop}%
\bibitem [{\citenamefont {Felder}\ and\ \citenamefont
  {Tkachev}(2008)}]{latticeeasy}%
  \BibitemOpen
  \bibfield  {author} {\bibinfo {author} {\bibfnamefont {G.}~\bibnamefont
  {Felder}}\ and\ \bibinfo {author} {\bibfnamefont {I.}~\bibnamefont
  {Tkachev}},\ }\href {https://doi.org/10.1016/j.cpc.2008.02.009} {\bibfield
  {journal} {\bibinfo  {journal} {Computer Physics Communications}\ }\textbf
  {\bibinfo {volume} {178}},\ \bibinfo {pages} {929?932} (\bibinfo {year}
  {2008})}\BibitemShut {NoStop}%
\bibitem [{\citenamefont {Garcia-Bellido}\ and\ \citenamefont
  {Figueroa}(2007)}]{Garcia-Bellido:2007nns}%
  \BibitemOpen
  \bibfield  {author} {\bibinfo {author} {\bibfnamefont {J.}~\bibnamefont
  {Garcia-Bellido}}\ and\ \bibinfo {author} {\bibfnamefont {D.~G.}\
  \bibnamefont {Figueroa}},\ }\href
  {https://doi.org/10.1103/PhysRevLett.98.061302} {\bibfield  {journal}
  {\bibinfo  {journal} {Phys. Rev. Lett.}\ }\textbf {\bibinfo {volume} {98}},\
  \bibinfo {pages} {061302} (\bibinfo {year} {2007})},\ \Eprint
  {https://arxiv.org/abs/astro-ph/0701014} {arXiv:astro-ph/0701014}
  \BibitemShut {NoStop}%
\bibitem [{\citenamefont {Frolov}(2008)}]{Frolov_2008}%
  \BibitemOpen
  \bibfield  {author} {\bibinfo {author} {\bibfnamefont {A.~V.}\ \bibnamefont
  {Frolov}},\ }\href {https://doi.org/10.1088/1475-7516/2008/11/009} {\bibfield
   {journal} {\bibinfo  {journal} {Journal of Cosmology and Astroparticle
  Physics}\ }\textbf {\bibinfo {volume} {2008}}\bibinfo  {number} { (11)},\
  \bibinfo {pages} {009}}\BibitemShut {NoStop}%
\bibitem [{\citenamefont {Huang}(2011)}]{hlattice}%
  \BibitemOpen
\bibfield  {number} {  }\bibfield  {author} {\bibinfo {author} {\bibfnamefont
  {Z.}~\bibnamefont {Huang}},\ }\bibfield  {journal} {\bibinfo  {journal}
  {Physical Review D}\ }\textbf {\bibinfo {volume} {83}},\ \href
  {https://doi.org/10.1103/physrevd.83.123509} {10.1103/physrevd.83.123509}
  (\bibinfo {year} {2011})\BibitemShut {NoStop}%
\bibitem [{\citenamefont {Sainio}(2012)}]{Sainio_2012}%
  \BibitemOpen
  \bibfield  {author} {\bibinfo {author} {\bibfnamefont {J.}~\bibnamefont
  {Sainio}},\ }\href {https://doi.org/10.1088/1475-7516/2012/04/038} {\bibfield
   {journal} {\bibinfo  {journal} {Journal of Cosmology and Astroparticle
  Physics}\ }\textbf {\bibinfo {volume} {2012}}\bibinfo  {number} { (04)},\
  \bibinfo {pages} {038?038}}\BibitemShut {NoStop}%
\bibitem [{\citenamefont {Child}\ \emph {et~al.}(2013)\citenamefont {Child},
  \citenamefont {Giblin}, \citenamefont {Ribeiro},\ and\ \citenamefont
  {Seery}}]{Child_2013}%
  \BibitemOpen
\bibfield  {number} {  }\bibfield  {author} {\bibinfo {author} {\bibfnamefont
  {H.~L.}\ \bibnamefont {Child}}, \bibinfo {author} {\bibfnamefont {J.~T.}\
  \bibnamefont {Giblin}}, \bibinfo {author} {\bibfnamefont {R.~H.}\
  \bibnamefont {Ribeiro}},\ and\ \bibinfo {author} {\bibfnamefont
  {D.}~\bibnamefont {Seery}},\ }\bibfield  {journal} {\bibinfo  {journal}
  {Physical Review Letters}\ }\textbf {\bibinfo {volume} {111}},\ \href
  {https://doi.org/10.1103/physrevlett.111.051301}
  {10.1103/physrevlett.111.051301} (\bibinfo {year} {2013})\BibitemShut
  {NoStop}%
\bibitem [{\citenamefont {Easther}\ \emph {et~al.}(2010)\citenamefont
  {Easther}, \citenamefont {Finkel},\ and\ \citenamefont
  {Roth}}]{Easther_2010}%
  \BibitemOpen
  \bibfield  {author} {\bibinfo {author} {\bibfnamefont {R.}~\bibnamefont
  {Easther}}, \bibinfo {author} {\bibfnamefont {H.}~\bibnamefont {Finkel}},\
  and\ \bibinfo {author} {\bibfnamefont {N.}~\bibnamefont {Roth}},\ }\href
  {https://doi.org/10.1088/1475-7516/2010/10/025} {\bibfield  {journal}
  {\bibinfo  {journal} {Journal of Cosmology and Astroparticle Physics}\
  }\textbf {\bibinfo {volume} {2010}}\bibinfo  {number} { (10)},\ \bibinfo
  {pages} {025?025}}\BibitemShut {NoStop}%
\bibitem [{\citenamefont {Lozanov}\ and\ \citenamefont
  {Amin}(2020)}]{Lozanov_2020}%
  \BibitemOpen
\bibfield  {number} {  }\bibfield  {author} {\bibinfo {author} {\bibfnamefont
  {K.~D.}\ \bibnamefont {Lozanov}}\ and\ \bibinfo {author} {\bibfnamefont
  {M.~A.}\ \bibnamefont {Amin}},\ }\href
  {https://doi.org/10.1088/1475-7516/2020/04/058} {\bibfield  {journal}
  {\bibinfo  {journal} {Journal of Cosmology and Astroparticle Physics}\
  }\textbf {\bibinfo {volume} {2020}}\bibinfo  {number} { (04)},\ \bibinfo
  {pages} {058?058}}\BibitemShut {NoStop}%
\bibitem [{\citenamefont {Figueroa}\ \emph {et~al.}(2021)\citenamefont
  {Figueroa}, \citenamefont {Florio}, \citenamefont {Torrenti},\ and\
  \citenamefont {Valkenburg}}]{figueroa2021cosmolattice}%
  \BibitemOpen
\bibfield  {number} {  }\bibfield  {author} {\bibinfo {author} {\bibfnamefont
  {D.~G.}\ \bibnamefont {Figueroa}}, \bibinfo {author} {\bibfnamefont
  {A.}~\bibnamefont {Florio}}, \bibinfo {author} {\bibfnamefont
  {F.}~\bibnamefont {Torrenti}},\ and\ \bibinfo {author} {\bibfnamefont
  {W.}~\bibnamefont {Valkenburg}},\ }\href@noop {} {\  (\bibinfo {year}
  {2021})},\ \Eprint {https://arxiv.org/abs/2102.01031} {arXiv:2102.01031
  [astro-ph.CO]} \BibitemShut {NoStop}%
\bibitem [{\citenamefont {Caravano}\ \emph {et~al.}(2021)\citenamefont
  {Caravano}, \citenamefont {Komatsu}, \citenamefont {Lozanov},\ and\
  \citenamefont {Weller}}]{Caravano:2021pgc}%
  \BibitemOpen
  \bibfield  {author} {\bibinfo {author} {\bibfnamefont {A.}~\bibnamefont
  {Caravano}}, \bibinfo {author} {\bibfnamefont {E.}~\bibnamefont {Komatsu}},
  \bibinfo {author} {\bibfnamefont {K.~D.}\ \bibnamefont {Lozanov}},\ and\
  \bibinfo {author} {\bibfnamefont {J.}~\bibnamefont {Weller}},\ }\href
  {https://doi.org/10.1088/1475-7516/2021/12/010} {\bibfield  {journal}
  {\bibinfo  {journal} {JCAP}\ }\textbf {\bibinfo {volume} {12}}\bibfield
  {number} {\bibinfo  {number} { (12)},\ \bibinfo {pages} {010}},\ }\Eprint
  {https://arxiv.org/abs/2102.06378} {arXiv:2102.06378 [astro-ph.CO]}
  \BibitemShut {NoStop}%
\bibitem [{\citenamefont {Caravano}\ \emph {et~al.}(2022)\citenamefont
  {Caravano}, \citenamefont {Komatsu}, \citenamefont {Lozanov},\ and\
  \citenamefont {Weller}}]{Caravano:2021bfn}%
  \BibitemOpen
  \bibfield  {author} {\bibinfo {author} {\bibfnamefont {A.}~\bibnamefont
  {Caravano}}, \bibinfo {author} {\bibfnamefont {E.}~\bibnamefont {Komatsu}},
  \bibinfo {author} {\bibfnamefont {K.~D.}\ \bibnamefont {Lozanov}},\ and\
  \bibinfo {author} {\bibfnamefont {J.}~\bibnamefont {Weller}},\ }\href
  {https://doi.org/10.1103/PhysRevD.105.123530} {\bibfield  {journal} {\bibinfo
   {journal} {Phys. Rev. D}\ }\textbf {\bibinfo {volume} {105}},\ \bibinfo
  {pages} {123530} (\bibinfo {year} {2022})},\ \Eprint
  {https://arxiv.org/abs/2110.10695} {arXiv:2110.10695 [astro-ph.CO]}
  \BibitemShut {NoStop}%
\bibitem [{\citenamefont {Caravano}\ \emph {et~al.}(2023)\citenamefont
  {Caravano}, \citenamefont {Komatsu}, \citenamefont {Lozanov},\ and\
  \citenamefont {Weller}}]{Caravano:2022epk}%
  \BibitemOpen
  \bibfield  {author} {\bibinfo {author} {\bibfnamefont {A.}~\bibnamefont
  {Caravano}}, \bibinfo {author} {\bibfnamefont {E.}~\bibnamefont {Komatsu}},
  \bibinfo {author} {\bibfnamefont {K.~D.}\ \bibnamefont {Lozanov}},\ and\
  \bibinfo {author} {\bibfnamefont {J.}~\bibnamefont {Weller}},\ }\href
  {https://doi.org/10.1103/PhysRevD.108.043504} {\bibfield  {journal} {\bibinfo
   {journal} {Phys. Rev. D}\ }\textbf {\bibinfo {volume} {108}},\ \bibinfo
  {pages} {043504} (\bibinfo {year} {2023})},\ \Eprint
  {https://arxiv.org/abs/2204.12874} {arXiv:2204.12874 [astro-ph.CO]}
  \BibitemShut {NoStop}%
\bibitem [{\citenamefont {Caravano}(2022)}]{Caravano:2022yyv}%
  \BibitemOpen
  \bibfield  {author} {\bibinfo {author} {\bibfnamefont {A.}~\bibnamefont
  {Caravano}},\ }\href {https://doi.org/10.5282/edoc.30905} {Ph.D. thesis},\
  \bibinfo  {school} {Munich U.} (\bibinfo {year} {2022}),\ \Eprint
  {https://arxiv.org/abs/2209.13616} {arXiv:2209.13616 [astro-ph.CO]}
  \BibitemShut {NoStop}%
\bibitem [{\citenamefont {Krajewski}\ and\ \citenamefont
  {Turzy\'nski}(2022)}]{Krajewski:2022uuh}%
  \BibitemOpen
  \bibfield  {author} {\bibinfo {author} {\bibfnamefont {T.}~\bibnamefont
  {Krajewski}}\ and\ \bibinfo {author} {\bibfnamefont {K.}~\bibnamefont
  {Turzy\'nski}},\ }\href {https://doi.org/10.1088/1475-7516/2022/10/064}
  {\bibfield  {journal} {\bibinfo  {journal} {JCAP}\ }\textbf {\bibinfo
  {volume} {10}},\ \bibinfo {pages} {064}},\ \Eprint
  {https://arxiv.org/abs/2205.13487} {arXiv:2205.13487 [astro-ph.CO]}
  \BibitemShut {NoStop}%
\bibitem [{\citenamefont {Figueroa}\ \emph {et~al.}(2023)\citenamefont
  {Figueroa}, \citenamefont {Lizarraga}, \citenamefont {Urio},\ and\
  \citenamefont {Urrestilla}}]{Figueroa:2023oxc}%
  \BibitemOpen
  \bibfield  {author} {\bibinfo {author} {\bibfnamefont {D.~G.}\ \bibnamefont
  {Figueroa}}, \bibinfo {author} {\bibfnamefont {J.}~\bibnamefont {Lizarraga}},
  \bibinfo {author} {\bibfnamefont {A.}~\bibnamefont {Urio}},\ and\ \bibinfo
  {author} {\bibfnamefont {J.}~\bibnamefont {Urrestilla}},\ }\href
  {https://doi.org/10.1103/PhysRevLett.131.151003} {\bibfield  {journal}
  {\bibinfo  {journal} {Phys. Rev. Lett.}\ }\textbf {\bibinfo {volume} {131}},\
  \bibinfo {pages} {151003} (\bibinfo {year} {2023})},\ \Eprint
  {https://arxiv.org/abs/2303.17436} {arXiv:2303.17436 [astro-ph.CO]}
  \BibitemShut {NoStop}%
\bibitem [{\citenamefont {East}\ \emph {et~al.}(2016)\citenamefont {East},
  \citenamefont {Kleban}, \citenamefont {Linde},\ and\ \citenamefont
  {Senatore}}]{East:2015ggf}%
  \BibitemOpen
  \bibfield  {author} {\bibinfo {author} {\bibfnamefont {W.~E.}\ \bibnamefont
  {East}}, \bibinfo {author} {\bibfnamefont {M.}~\bibnamefont {Kleban}},
  \bibinfo {author} {\bibfnamefont {A.}~\bibnamefont {Linde}},\ and\ \bibinfo
  {author} {\bibfnamefont {L.}~\bibnamefont {Senatore}},\ }\href
  {https://doi.org/10.1088/1475-7516/2016/09/010} {\bibfield  {journal}
  {\bibinfo  {journal} {JCAP}\ }\textbf {\bibinfo {volume} {09}},\ \bibinfo
  {pages} {010}},\ \Eprint {https://arxiv.org/abs/1511.05143} {arXiv:1511.05143
  [hep-th]} \BibitemShut {NoStop}%
\bibitem [{\citenamefont {Braden}\ \emph {et~al.}(2017)\citenamefont {Braden},
  \citenamefont {Johnson}, \citenamefont {Peiris},\ and\ \citenamefont
  {Aguirre}}]{Braden:2016tjn}%
  \BibitemOpen
  \bibfield  {author} {\bibinfo {author} {\bibfnamefont {J.}~\bibnamefont
  {Braden}}, \bibinfo {author} {\bibfnamefont {M.~C.}\ \bibnamefont {Johnson}},
  \bibinfo {author} {\bibfnamefont {H.~V.}\ \bibnamefont {Peiris}},\ and\
  \bibinfo {author} {\bibfnamefont {A.}~\bibnamefont {Aguirre}},\ }\href
  {https://doi.org/10.1103/PhysRevD.96.023541} {\bibfield  {journal} {\bibinfo
  {journal} {Phys. Rev. D}\ }\textbf {\bibinfo {volume} {96}},\ \bibinfo
  {pages} {023541} (\bibinfo {year} {2017})},\ \Eprint
  {https://arxiv.org/abs/1604.04001} {arXiv:1604.04001 [astro-ph.CO]}
  \BibitemShut {NoStop}%
\bibitem [{\citenamefont {Clough}\ \emph {et~al.}(2017)\citenamefont {Clough},
  \citenamefont {Lim}, \citenamefont {DiNunno}, \citenamefont {Fischler},
  \citenamefont {Flauger},\ and\ \citenamefont {Paban}}]{Clough:2016ymm}%
  \BibitemOpen
  \bibfield  {author} {\bibinfo {author} {\bibfnamefont {K.}~\bibnamefont
  {Clough}}, \bibinfo {author} {\bibfnamefont {E.~A.}\ \bibnamefont {Lim}},
  \bibinfo {author} {\bibfnamefont {B.~S.}\ \bibnamefont {DiNunno}}, \bibinfo
  {author} {\bibfnamefont {W.}~\bibnamefont {Fischler}}, \bibinfo {author}
  {\bibfnamefont {R.}~\bibnamefont {Flauger}},\ and\ \bibinfo {author}
  {\bibfnamefont {S.}~\bibnamefont {Paban}},\ }\href
  {https://doi.org/10.1088/1475-7516/2017/09/025} {\bibfield  {journal}
  {\bibinfo  {journal} {JCAP}\ }\textbf {\bibinfo {volume} {09}},\ \bibinfo
  {pages} {025}},\ \Eprint {https://arxiv.org/abs/1608.04408} {arXiv:1608.04408
  [hep-th]} \BibitemShut {NoStop}%
\bibitem [{\citenamefont {Clough}\ \emph {et~al.}(2018)\citenamefont {Clough},
  \citenamefont {Flauger},\ and\ \citenamefont {Lim}}]{Clough:2017efm}%
  \BibitemOpen
  \bibfield  {author} {\bibinfo {author} {\bibfnamefont {K.}~\bibnamefont
  {Clough}}, \bibinfo {author} {\bibfnamefont {R.}~\bibnamefont {Flauger}},\
  and\ \bibinfo {author} {\bibfnamefont {E.~A.}\ \bibnamefont {Lim}},\ }\href
  {https://doi.org/10.1088/1475-7516/2018/05/065} {\bibfield  {journal}
  {\bibinfo  {journal} {JCAP}\ }\textbf {\bibinfo {volume} {05}},\ \bibinfo
  {pages} {065}},\ \Eprint {https://arxiv.org/abs/1712.07352} {arXiv:1712.07352
  [hep-th]} \BibitemShut {NoStop}%
\bibitem [{\citenamefont {Chen}\ \emph {et~al.}(2008)\citenamefont {Chen},
  \citenamefont {Easther},\ and\ \citenamefont {Lim}}]{Chen:2008wn}%
  \BibitemOpen
  \bibfield  {author} {\bibinfo {author} {\bibfnamefont {X.}~\bibnamefont
  {Chen}}, \bibinfo {author} {\bibfnamefont {R.}~\bibnamefont {Easther}},\ and\
  \bibinfo {author} {\bibfnamefont {E.~A.}\ \bibnamefont {Lim}},\ }\href
  {https://doi.org/10.1088/1475-7516/2008/04/010} {\bibfield  {journal}
  {\bibinfo  {journal} {JCAP}\ }\textbf {\bibinfo {volume} {04}},\ \bibinfo
  {pages} {010}},\ \Eprint {https://arxiv.org/abs/0801.3295} {arXiv:0801.3295
  [astro-ph]} \BibitemShut {NoStop}%
\bibitem [{\citenamefont {Silverstein}\ and\ \citenamefont
  {Westphal}(2008)}]{Silverstein:2008sg}%
  \BibitemOpen
  \bibfield  {author} {\bibinfo {author} {\bibfnamefont {E.}~\bibnamefont
  {Silverstein}}\ and\ \bibinfo {author} {\bibfnamefont {A.}~\bibnamefont
  {Westphal}},\ }\href {https://doi.org/10.1103/PhysRevD.78.106003} {\bibfield
  {journal} {\bibinfo  {journal} {Phys. Rev. D}\ }\textbf {\bibinfo {volume}
  {78}},\ \bibinfo {pages} {106003} (\bibinfo {year} {2008})},\ \Eprint
  {https://arxiv.org/abs/0803.3085} {arXiv:0803.3085 [hep-th]} \BibitemShut
  {NoStop}%
\bibitem [{\citenamefont {McAllister}\ \emph {et~al.}(2010)\citenamefont
  {McAllister}, \citenamefont {Silverstein},\ and\ \citenamefont
  {Westphal}}]{McAllister:2008hb}%
  \BibitemOpen
  \bibfield  {author} {\bibinfo {author} {\bibfnamefont {L.}~\bibnamefont
  {McAllister}}, \bibinfo {author} {\bibfnamefont {E.}~\bibnamefont
  {Silverstein}},\ and\ \bibinfo {author} {\bibfnamefont {A.}~\bibnamefont
  {Westphal}},\ }\href {https://doi.org/10.1103/PhysRevD.82.046003} {\bibfield
  {journal} {\bibinfo  {journal} {Phys. Rev. D}\ }\textbf {\bibinfo {volume}
  {82}},\ \bibinfo {pages} {046003} (\bibinfo {year} {2010})},\ \Eprint
  {https://arxiv.org/abs/0808.0706} {arXiv:0808.0706 [hep-th]} \BibitemShut
  {NoStop}%
\bibitem [{\citenamefont {Flauger}\ \emph {et~al.}(2010)\citenamefont
  {Flauger}, \citenamefont {McAllister}, \citenamefont {Pajer}, \citenamefont
  {Westphal},\ and\ \citenamefont {Xu}}]{Flauger:2009ab}%
  \BibitemOpen
  \bibfield  {author} {\bibinfo {author} {\bibfnamefont {R.}~\bibnamefont
  {Flauger}}, \bibinfo {author} {\bibfnamefont {L.}~\bibnamefont {McAllister}},
  \bibinfo {author} {\bibfnamefont {E.}~\bibnamefont {Pajer}}, \bibinfo
  {author} {\bibfnamefont {A.}~\bibnamefont {Westphal}},\ and\ \bibinfo
  {author} {\bibfnamefont {G.}~\bibnamefont {Xu}},\ }\href
  {https://doi.org/10.1088/1475-7516/2010/06/009} {\bibfield  {journal}
  {\bibinfo  {journal} {JCAP}\ }\textbf {\bibinfo {volume} {06}},\ \bibinfo
  {pages} {009}},\ \Eprint {https://arxiv.org/abs/0907.2916} {arXiv:0907.2916
  [hep-th]} \BibitemShut {NoStop}%
\bibitem [{\citenamefont {Creminelli}\ \emph {et~al.}(2024)\citenamefont
  {Creminelli}, \citenamefont {Renaux-Petel}, \citenamefont {Tambalo},\ and\
  \citenamefont {Yingcharoenrat}}]{Creminelli:2024cge}%
  \BibitemOpen
  \bibfield  {author} {\bibinfo {author} {\bibfnamefont {P.}~\bibnamefont
  {Creminelli}}, \bibinfo {author} {\bibfnamefont {S.}~\bibnamefont
  {Renaux-Petel}}, \bibinfo {author} {\bibfnamefont {G.}~\bibnamefont
  {Tambalo}},\ and\ \bibinfo {author} {\bibfnamefont {V.}~\bibnamefont
  {Yingcharoenrat}},\ }\href@noop {} {\  (\bibinfo {year} {2024})},\ \Eprint
  {https://arxiv.org/abs/2401.10212} {arXiv:2401.10212 [hep-th]} \BibitemShut
  {NoStop}%
\bibitem [{\citenamefont {Flauger}\ and\ \citenamefont
  {Pajer}(2011)}]{Flauger:2010ja}%
  \BibitemOpen
  \bibfield  {author} {\bibinfo {author} {\bibfnamefont {R.}~\bibnamefont
  {Flauger}}\ and\ \bibinfo {author} {\bibfnamefont {E.}~\bibnamefont
  {Pajer}},\ }\href {https://doi.org/10.1088/1475-7516/2011/01/017} {\bibfield
  {journal} {\bibinfo  {journal} {JCAP}\ }\textbf {\bibinfo {volume} {01}},\
  \bibinfo {pages} {017}},\ \Eprint {https://arxiv.org/abs/1002.0833}
  {arXiv:1002.0833 [hep-th]} \BibitemShut {NoStop}%
\bibitem [{\citenamefont {Behbahani}\ \emph {et~al.}(2012)\citenamefont
  {Behbahani}, \citenamefont {Dymarsky}, \citenamefont {Mirbabayi},\ and\
  \citenamefont {Senatore}}]{Behbahani:2011it}%
  \BibitemOpen
  \bibfield  {author} {\bibinfo {author} {\bibfnamefont {S.~R.}\ \bibnamefont
  {Behbahani}}, \bibinfo {author} {\bibfnamefont {A.}~\bibnamefont {Dymarsky}},
  \bibinfo {author} {\bibfnamefont {M.}~\bibnamefont {Mirbabayi}},\ and\
  \bibinfo {author} {\bibfnamefont {L.}~\bibnamefont {Senatore}},\ }\href
  {https://doi.org/10.1088/1475-7516/2012/12/036} {\bibfield  {journal}
  {\bibinfo  {journal} {JCAP}\ }\textbf {\bibinfo {volume} {12}},\ \bibinfo
  {pages} {036}},\ \Eprint {https://arxiv.org/abs/1111.3373} {arXiv:1111.3373
  [hep-th]} \BibitemShut {NoStop}%
\bibitem [{\citenamefont {Duaso~Pueyo}\ and\ \citenamefont
  {Pajer}(2023)}]{DuasoPueyo:2023viy}%
  \BibitemOpen
  \bibfield  {author} {\bibinfo {author} {\bibfnamefont {C.}~\bibnamefont
  {Duaso~Pueyo}}\ and\ \bibinfo {author} {\bibfnamefont {E.}~\bibnamefont
  {Pajer}},\ }\href@noop {} {\  (\bibinfo {year} {2023})},\ \Eprint
  {https://arxiv.org/abs/2311.01395} {arXiv:2311.01395 [hep-th]} \BibitemShut
  {NoStop}%
\bibitem [{\citenamefont {Sasaki}\ \emph {et~al.}(2018)\citenamefont {Sasaki},
  \citenamefont {Suyama}, \citenamefont {Tanaka},\ and\ \citenamefont
  {Yokoyama}}]{Sasaki:2018dmp}%
  \BibitemOpen
  \bibfield  {author} {\bibinfo {author} {\bibfnamefont {M.}~\bibnamefont
  {Sasaki}}, \bibinfo {author} {\bibfnamefont {T.}~\bibnamefont {Suyama}},
  \bibinfo {author} {\bibfnamefont {T.}~\bibnamefont {Tanaka}},\ and\ \bibinfo
  {author} {\bibfnamefont {S.}~\bibnamefont {Yokoyama}},\ }\href
  {https://doi.org/10.1088/1361-6382/aaa7b4} {\bibfield  {journal} {\bibinfo
  {journal} {Class. Quant. Grav.}\ }\textbf {\bibinfo {volume} {35}},\ \bibinfo
  {pages} {063001} (\bibinfo {year} {2018})},\ \Eprint
  {https://arxiv.org/abs/1801.05235} {arXiv:1801.05235 [astro-ph.CO]}
  \BibitemShut {NoStop}%
\bibitem [{\citenamefont {Cheung}\ \emph {et~al.}(2008)\citenamefont {Cheung},
  \citenamefont {Creminelli}, \citenamefont {Fitzpatrick}, \citenamefont
  {Kaplan},\ and\ \citenamefont {Senatore}}]{Cheung:2007st}%
  \BibitemOpen
  \bibfield  {author} {\bibinfo {author} {\bibfnamefont {C.}~\bibnamefont
  {Cheung}}, \bibinfo {author} {\bibfnamefont {P.}~\bibnamefont {Creminelli}},
  \bibinfo {author} {\bibfnamefont {A.~L.}\ \bibnamefont {Fitzpatrick}},
  \bibinfo {author} {\bibfnamefont {J.}~\bibnamefont {Kaplan}},\ and\ \bibinfo
  {author} {\bibfnamefont {L.}~\bibnamefont {Senatore}},\ }\href
  {https://doi.org/10.1088/1126-6708/2008/03/014} {\bibfield  {journal}
  {\bibinfo  {journal} {JHEP}\ }\textbf {\bibinfo {volume} {03}},\ \bibinfo
  {pages} {014}},\ \Eprint {https://arxiv.org/abs/0709.0293} {arXiv:0709.0293
  [hep-th]} \BibitemShut {NoStop}%
\bibitem [{\citenamefont {Armendariz-Picon}(2019)}]{Armendariz-Picon:2019csc}%
  \BibitemOpen
  \bibfield  {author} {\bibinfo {author} {\bibfnamefont {C.}~\bibnamefont
  {Armendariz-Picon}},\ }\href {https://doi.org/10.1088/1475-7516/2019/08/012}
  {\bibfield  {journal} {\bibinfo  {journal} {JCAP}\ }\textbf {\bibinfo
  {volume} {08}},\ \bibinfo {pages} {012}},\ \Eprint
  {https://arxiv.org/abs/1905.05724} {arXiv:1905.05724 [astro-ph.CO]}
  \BibitemShut {NoStop}%
\bibitem [{\citenamefont {Fumagalli}\ \emph {et~al.}(2023)\citenamefont
  {Fumagalli}, \citenamefont {Bhattacharya}, \citenamefont {Peloso},
  \citenamefont {Renaux-Petel},\ and\ \citenamefont
  {Witkowski}}]{Fumagalli:2023loc}%
  \BibitemOpen
  \bibfield  {author} {\bibinfo {author} {\bibfnamefont {J.}~\bibnamefont
  {Fumagalli}}, \bibinfo {author} {\bibfnamefont {S.}~\bibnamefont
  {Bhattacharya}}, \bibinfo {author} {\bibfnamefont {M.}~\bibnamefont
  {Peloso}}, \bibinfo {author} {\bibfnamefont {S.}~\bibnamefont
  {Renaux-Petel}},\ and\ \bibinfo {author} {\bibfnamefont {L.~T.}\ \bibnamefont
  {Witkowski}},\ }\href@noop {} {\  (\bibinfo {year} {2023})},\ \Eprint
  {https://arxiv.org/abs/2307.08358} {arXiv:2307.08358 [astro-ph.CO]}
  \BibitemShut {NoStop}%
\bibitem [{\citenamefont {Braden}\ \emph {et~al.}(2019)\citenamefont {Braden},
  \citenamefont {Johnson}, \citenamefont {Peiris}, \citenamefont {Pontzen},\
  and\ \citenamefont {Weinfurtner}}]{Braden:2018tky}%
  \BibitemOpen
  \bibfield  {author} {\bibinfo {author} {\bibfnamefont {J.}~\bibnamefont
  {Braden}}, \bibinfo {author} {\bibfnamefont {M.~C.}\ \bibnamefont {Johnson}},
  \bibinfo {author} {\bibfnamefont {H.~V.}\ \bibnamefont {Peiris}}, \bibinfo
  {author} {\bibfnamefont {A.}~\bibnamefont {Pontzen}},\ and\ \bibinfo {author}
  {\bibfnamefont {S.}~\bibnamefont {Weinfurtner}},\ }\href
  {https://doi.org/10.1103/PhysRevLett.123.031601} {\bibfield  {journal}
  {\bibinfo  {journal} {Phys. Rev. Lett.}\ }\textbf {\bibinfo {volume} {123}},\
  \bibinfo {pages} {031601} (\bibinfo {year} {2019})},\ \bibinfo {note}
  {[Erratum: Phys.Rev.Lett. 129, 059901 (2022)]},\ \Eprint
  {https://arxiv.org/abs/1806.06069} {arXiv:1806.06069 [hep-th]} \BibitemShut
  {NoStop}%
\bibitem [{\citenamefont {Hertzberg}\ and\ \citenamefont
  {Yamada}(2019)}]{Hertzberg:2019wgx}%
  \BibitemOpen
  \bibfield  {author} {\bibinfo {author} {\bibfnamefont {M.~P.}\ \bibnamefont
  {Hertzberg}}\ and\ \bibinfo {author} {\bibfnamefont {M.}~\bibnamefont
  {Yamada}},\ }\href {https://doi.org/10.1103/PhysRevD.100.016011} {\bibfield
  {journal} {\bibinfo  {journal} {Phys. Rev. D}\ }\textbf {\bibinfo {volume}
  {100}},\ \bibinfo {pages} {016011} (\bibinfo {year} {2019})},\ \Eprint
  {https://arxiv.org/abs/1904.08565} {arXiv:1904.08565 [hep-th]} \BibitemShut
  {NoStop}%
\bibitem [{\citenamefont {Hertzberg}\ \emph {et~al.}(2020)\citenamefont
  {Hertzberg}, \citenamefont {Rompineve},\ and\ \citenamefont
  {Shah}}]{Hertzberg:2020tqa}%
  \BibitemOpen
  \bibfield  {author} {\bibinfo {author} {\bibfnamefont {M.~P.}\ \bibnamefont
  {Hertzberg}}, \bibinfo {author} {\bibfnamefont {F.}~\bibnamefont
  {Rompineve}},\ and\ \bibinfo {author} {\bibfnamefont {N.}~\bibnamefont
  {Shah}},\ }\href {https://doi.org/10.1103/PhysRevD.102.076003} {\bibfield
  {journal} {\bibinfo  {journal} {Phys. Rev. D}\ }\textbf {\bibinfo {volume}
  {102}},\ \bibinfo {pages} {076003} (\bibinfo {year} {2020})},\ \Eprint
  {https://arxiv.org/abs/2009.00017} {arXiv:2009.00017 [hep-th]} \BibitemShut
  {NoStop}%
\bibitem [{\citenamefont {Blanco-Pillado}\ \emph {et~al.}(2019)\citenamefont
  {Blanco-Pillado}, \citenamefont {Deng},\ and\ \citenamefont
  {Vilenkin}}]{Blanco-Pillado:2019xny}%
  \BibitemOpen
  \bibfield  {author} {\bibinfo {author} {\bibfnamefont {J.~J.}\ \bibnamefont
  {Blanco-Pillado}}, \bibinfo {author} {\bibfnamefont {H.}~\bibnamefont
  {Deng}},\ and\ \bibinfo {author} {\bibfnamefont {A.}~\bibnamefont
  {Vilenkin}},\ }\href {https://doi.org/10.1088/1475-7516/2019/12/001}
  {\bibfield  {journal} {\bibinfo  {journal} {JCAP}\ }\textbf {\bibinfo
  {volume} {12}},\ \bibinfo {pages} {001}},\ \Eprint
  {https://arxiv.org/abs/1906.09657} {arXiv:1906.09657 [hep-th]} \BibitemShut
  {NoStop}%
\bibitem [{\citenamefont {Mou}\ \emph {et~al.}(2019)\citenamefont {Mou},
  \citenamefont {Saffin},\ and\ \citenamefont {Tranberg}}]{Mou:2019gyl}%
  \BibitemOpen
  \bibfield  {author} {\bibinfo {author} {\bibfnamefont {Z.-G.}\ \bibnamefont
  {Mou}}, \bibinfo {author} {\bibfnamefont {P.~M.}\ \bibnamefont {Saffin}},\
  and\ \bibinfo {author} {\bibfnamefont {A.}~\bibnamefont {Tranberg}},\ }\href
  {https://doi.org/10.1007/JHEP11(2019)135} {\bibfield  {journal} {\bibinfo
  {journal} {JHEP}\ }\textbf {\bibinfo {volume} {11}},\ \bibinfo {pages}
  {135}},\ \Eprint {https://arxiv.org/abs/1909.02488} {arXiv:1909.02488
  [hep-th]} \BibitemShut {NoStop}%
\bibitem [{\citenamefont {Ai}\ \emph {et~al.}(2019)\citenamefont {Ai},
  \citenamefont {Garbrecht},\ and\ \citenamefont {Tamarit}}]{Ai:2019fri}%
  \BibitemOpen
  \bibfield  {author} {\bibinfo {author} {\bibfnamefont {W.-Y.}\ \bibnamefont
  {Ai}}, \bibinfo {author} {\bibfnamefont {B.}~\bibnamefont {Garbrecht}},\ and\
  \bibinfo {author} {\bibfnamefont {C.}~\bibnamefont {Tamarit}},\ }\href
  {https://doi.org/10.1007/JHEP12(2019)095} {\bibfield  {journal} {\bibinfo
  {journal} {JHEP}\ }\textbf {\bibinfo {volume} {12}},\ \bibinfo {pages}
  {095}},\ \Eprint {https://arxiv.org/abs/1905.04236} {arXiv:1905.04236
  [hep-th]} \BibitemShut {NoStop}%
\bibitem [{\citenamefont {Michel}(2020)}]{Michel:2019nwa}%
  \BibitemOpen
  \bibfield  {author} {\bibinfo {author} {\bibfnamefont {F.}~\bibnamefont
  {Michel}},\ }\href {https://doi.org/10.1103/PhysRevD.101.045021} {\bibfield
  {journal} {\bibinfo  {journal} {Phys. Rev. D}\ }\textbf {\bibinfo {volume}
  {101}},\ \bibinfo {pages} {045021} (\bibinfo {year} {2020})},\ \Eprint
  {https://arxiv.org/abs/1911.12765} {arXiv:1911.12765 [quant-ph]} \BibitemShut
  {NoStop}%
\bibitem [{\citenamefont {Huang}\ and\ \citenamefont
  {Ford}(2022)}]{Huang:2020bzb}%
  \BibitemOpen
  \bibfield  {author} {\bibinfo {author} {\bibfnamefont {H.}~\bibnamefont
  {Huang}}\ and\ \bibinfo {author} {\bibfnamefont {L.~H.}\ \bibnamefont
  {Ford}},\ }\href {https://doi.org/10.1103/PhysRevD.105.085025} {\bibfield
  {journal} {\bibinfo  {journal} {Phys. Rev. D}\ }\textbf {\bibinfo {volume}
  {105}},\ \bibinfo {pages} {085025} (\bibinfo {year} {2022})},\ \Eprint
  {https://arxiv.org/abs/2005.08355} {arXiv:2005.08355 [hep-th]} \BibitemShut
  {NoStop}%
\bibitem [{\citenamefont {Pirvu}\ \emph {et~al.}(2022)\citenamefont {Pirvu},
  \citenamefont {Braden},\ and\ \citenamefont {Johnson}}]{Pirvu:2021roq}%
  \BibitemOpen
  \bibfield  {author} {\bibinfo {author} {\bibfnamefont {D.}~\bibnamefont
  {Pirvu}}, \bibinfo {author} {\bibfnamefont {J.}~\bibnamefont {Braden}},\ and\
  \bibinfo {author} {\bibfnamefont {M.~C.}\ \bibnamefont {Johnson}},\ }\href
  {https://doi.org/10.1103/PhysRevD.105.043510} {\bibfield  {journal} {\bibinfo
   {journal} {Phys. Rev. D}\ }\textbf {\bibinfo {volume} {105}},\ \bibinfo
  {pages} {043510} (\bibinfo {year} {2022})},\ \Eprint
  {https://arxiv.org/abs/2109.04496} {arXiv:2109.04496 [hep-th]} \BibitemShut
  {NoStop}%
\bibitem [{\citenamefont {Braden}\ \emph {et~al.}(2023)\citenamefont {Braden},
  \citenamefont {Johnson}, \citenamefont {Peiris}, \citenamefont {Pontzen},\
  and\ \citenamefont {Weinfurtner}}]{Braden:2022odm}%
  \BibitemOpen
  \bibfield  {author} {\bibinfo {author} {\bibfnamefont {J.}~\bibnamefont
  {Braden}}, \bibinfo {author} {\bibfnamefont {M.~C.}\ \bibnamefont {Johnson}},
  \bibinfo {author} {\bibfnamefont {H.~V.}\ \bibnamefont {Peiris}}, \bibinfo
  {author} {\bibfnamefont {A.}~\bibnamefont {Pontzen}},\ and\ \bibinfo {author}
  {\bibfnamefont {S.}~\bibnamefont {Weinfurtner}},\ }\href
  {https://doi.org/10.1103/PhysRevD.107.083509} {\bibfield  {journal} {\bibinfo
   {journal} {Phys. Rev. D}\ }\textbf {\bibinfo {volume} {107}},\ \bibinfo
  {pages} {083509} (\bibinfo {year} {2023})},\ \Eprint
  {https://arxiv.org/abs/2204.11867} {arXiv:2204.11867 [hep-th]} \BibitemShut
  {NoStop}%
\bibitem [{\citenamefont {Batini}\ \emph {et~al.}(2023)\citenamefont {Batini},
  \citenamefont {Chatrchyan},\ and\ \citenamefont {Berges}}]{Batini:2023zpi}%
  \BibitemOpen
  \bibfield  {author} {\bibinfo {author} {\bibfnamefont {L.}~\bibnamefont
  {Batini}}, \bibinfo {author} {\bibfnamefont {A.}~\bibnamefont {Chatrchyan}},\
  and\ \bibinfo {author} {\bibfnamefont {J.}~\bibnamefont {Berges}},\
  }\href@noop {} {\  (\bibinfo {year} {2023})},\ \Eprint
  {https://arxiv.org/abs/2310.04206} {arXiv:2310.04206 [hep-th]} \BibitemShut
  {NoStop}%
\bibitem [{\citenamefont {Ai}\ \emph {et~al.}(2023)\citenamefont {Ai},
  \citenamefont {Alexandre},\ and\ \citenamefont {Sarkar}}]{Ai:2023yce}%
  \BibitemOpen
  \bibfield  {author} {\bibinfo {author} {\bibfnamefont {W.-Y.}\ \bibnamefont
  {Ai}}, \bibinfo {author} {\bibfnamefont {J.}~\bibnamefont {Alexandre}},\ and\
  \bibinfo {author} {\bibfnamefont {S.}~\bibnamefont {Sarkar}},\ }\href@noop {}
  {\  (\bibinfo {year} {2023})},\ \Eprint {https://arxiv.org/abs/2312.04482}
  {arXiv:2312.04482 [hep-ph]} \BibitemShut {NoStop}%
\bibitem [{\citenamefont {Miyachi}\ \emph {et~al.}(2023)\citenamefont
  {Miyachi}, \citenamefont {Soda},\ and\ \citenamefont
  {Tokuda}}]{Miyachi:2023fss}%
  \BibitemOpen
  \bibfield  {author} {\bibinfo {author} {\bibfnamefont {T.}~\bibnamefont
  {Miyachi}}, \bibinfo {author} {\bibfnamefont {J.}~\bibnamefont {Soda}},\ and\
  \bibinfo {author} {\bibfnamefont {J.}~\bibnamefont {Tokuda}},\ }\href@noop {}
  {\  (\bibinfo {year} {2023})},\ \Eprint {https://arxiv.org/abs/2309.07440}
  {arXiv:2309.07440 [hep-th]} \BibitemShut {NoStop}%
\bibitem [{\citenamefont {Deng}\ and\ \citenamefont
  {Vilenkin}(2017)}]{Deng:2017uwc}%
  \BibitemOpen
  \bibfield  {author} {\bibinfo {author} {\bibfnamefont {H.}~\bibnamefont
  {Deng}}\ and\ \bibinfo {author} {\bibfnamefont {A.}~\bibnamefont
  {Vilenkin}},\ }\href {https://doi.org/10.1088/1475-7516/2017/12/044}
  {\bibfield  {journal} {\bibinfo  {journal} {JCAP}\ }\textbf {\bibinfo
  {volume} {12}},\ \bibinfo {pages} {044}},\ \Eprint
  {https://arxiv.org/abs/1710.02865} {arXiv:1710.02865 [gr-qc]} \BibitemShut
  {NoStop}%
\bibitem [{\citenamefont {Garriga}\ \emph {et~al.}(2016)\citenamefont
  {Garriga}, \citenamefont {Vilenkin},\ and\ \citenamefont
  {Zhang}}]{Garriga:2015fdk}%
  \BibitemOpen
  \bibfield  {author} {\bibinfo {author} {\bibfnamefont {J.}~\bibnamefont
  {Garriga}}, \bibinfo {author} {\bibfnamefont {A.}~\bibnamefont {Vilenkin}},\
  and\ \bibinfo {author} {\bibfnamefont {J.}~\bibnamefont {Zhang}},\ }\href
  {https://doi.org/10.1088/1475-7516/2016/02/064} {\bibfield  {journal}
  {\bibinfo  {journal} {JCAP}\ }\textbf {\bibinfo {volume} {02}},\ \bibinfo
  {pages} {064}},\ \Eprint {https://arxiv.org/abs/1512.01819} {arXiv:1512.01819
  [hep-th]} \BibitemShut {NoStop}%
\bibitem [{\citenamefont {Atal}\ \emph {et~al.}(2019)\citenamefont {Atal},
  \citenamefont {Garriga},\ and\ \citenamefont
  {Marcos-Caballero}}]{Atal:2019cdz}%
  \BibitemOpen
  \bibfield  {author} {\bibinfo {author} {\bibfnamefont {V.}~\bibnamefont
  {Atal}}, \bibinfo {author} {\bibfnamefont {J.}~\bibnamefont {Garriga}},\ and\
  \bibinfo {author} {\bibfnamefont {A.}~\bibnamefont {Marcos-Caballero}},\
  }\href {https://doi.org/10.1088/1475-7516/2019/09/073} {\bibfield  {journal}
  {\bibinfo  {journal} {JCAP}\ }\textbf {\bibinfo {volume} {09}},\ \bibinfo
  {pages} {073}},\ \Eprint {https://arxiv.org/abs/1905.13202} {arXiv:1905.13202
  [astro-ph.CO]} \BibitemShut {NoStop}%
\bibitem [{\citenamefont {Atal}\ \emph {et~al.}(2020)\citenamefont {Atal},
  \citenamefont {Cid}, \citenamefont {Escriv\`a},\ and\ \citenamefont
  {Garriga}}]{Atal:2019erb}%
  \BibitemOpen
  \bibfield  {author} {\bibinfo {author} {\bibfnamefont {V.}~\bibnamefont
  {Atal}}, \bibinfo {author} {\bibfnamefont {J.}~\bibnamefont {Cid}}, \bibinfo
  {author} {\bibfnamefont {A.}~\bibnamefont {Escriv\`a}},\ and\ \bibinfo
  {author} {\bibfnamefont {J.}~\bibnamefont {Garriga}},\ }\href
  {https://doi.org/10.1088/1475-7516/2020/05/022} {\bibfield  {journal}
  {\bibinfo  {journal} {JCAP}\ }\textbf {\bibinfo {volume} {05}},\ \bibinfo
  {pages} {022}},\ \Eprint {https://arxiv.org/abs/1908.11357} {arXiv:1908.11357
  [astro-ph.CO]} \BibitemShut {NoStop}%
\bibitem [{\citenamefont {Escriv\`a}\ \emph {et~al.}(2023)\citenamefont
  {Escriv\`a}, \citenamefont {Atal},\ and\ \citenamefont
  {Garriga}}]{Escriva:2023uko}%
  \BibitemOpen
  \bibfield  {author} {\bibinfo {author} {\bibfnamefont {A.}~\bibnamefont
  {Escriv\`a}}, \bibinfo {author} {\bibfnamefont {V.}~\bibnamefont {Atal}},\
  and\ \bibinfo {author} {\bibfnamefont {J.}~\bibnamefont {Garriga}},\ }\href
  {https://doi.org/10.1088/1475-7516/2023/10/035} {\bibfield  {journal}
  {\bibinfo  {journal} {JCAP}\ }\textbf {\bibinfo {volume} {10}},\ \bibinfo
  {pages} {035}},\ \Eprint {https://arxiv.org/abs/2306.09990} {arXiv:2306.09990
  [astro-ph.CO]} \BibitemShut {NoStop}%
\bibitem [{\citenamefont {Huang}\ \emph {et~al.}(2023)\citenamefont {Huang},
  \citenamefont {Cai}, \citenamefont {Jiang}, \citenamefont {Zhang},\ and\
  \citenamefont {Piao}}]{Huang:2023chx}%
  \BibitemOpen
  \bibfield  {author} {\bibinfo {author} {\bibfnamefont {H.-L.}\ \bibnamefont
  {Huang}}, \bibinfo {author} {\bibfnamefont {Y.}~\bibnamefont {Cai}}, \bibinfo
  {author} {\bibfnamefont {J.-Q.}\ \bibnamefont {Jiang}}, \bibinfo {author}
  {\bibfnamefont {J.}~\bibnamefont {Zhang}},\ and\ \bibinfo {author}
  {\bibfnamefont {Y.-S.}\ \bibnamefont {Piao}},\ }\href@noop {} {\  (\bibinfo
  {year} {2023})},\ \Eprint {https://arxiv.org/abs/2306.17577}
  {arXiv:2306.17577 [gr-qc]} \BibitemShut {NoStop}%
\bibitem [{\citenamefont {Imrith}\ \emph {et~al.}(2018)\citenamefont {Imrith},
  \citenamefont {Mulryne},\ and\ \citenamefont {Rajantie}}]{Imrith:2018uyk}%
  \BibitemOpen
  \bibfield  {author} {\bibinfo {author} {\bibfnamefont {S.~V.}\ \bibnamefont
  {Imrith}}, \bibinfo {author} {\bibfnamefont {D.~J.}\ \bibnamefont
  {Mulryne}},\ and\ \bibinfo {author} {\bibfnamefont {A.}~\bibnamefont
  {Rajantie}},\ }\href {https://doi.org/10.1103/PhysRevD.98.043513} {\bibfield
  {journal} {\bibinfo  {journal} {Phys. Rev. D}\ }\textbf {\bibinfo {volume}
  {98}},\ \bibinfo {pages} {043513} (\bibinfo {year} {2018})},\ \Eprint
  {https://arxiv.org/abs/1801.02600} {arXiv:1801.02600 [astro-ph.CO]}
  \BibitemShut {NoStop}%
\bibitem [{\citenamefont {Imrith}\ \emph {et~al.}(2019)\citenamefont {Imrith},
  \citenamefont {Mulryne},\ and\ \citenamefont {Rajantie}}]{Imrith:2019njf}%
  \BibitemOpen
  \bibfield  {author} {\bibinfo {author} {\bibfnamefont {S.~V.}\ \bibnamefont
  {Imrith}}, \bibinfo {author} {\bibfnamefont {D.~J.}\ \bibnamefont
  {Mulryne}},\ and\ \bibinfo {author} {\bibfnamefont {A.}~\bibnamefont
  {Rajantie}},\ }\href {https://doi.org/10.1103/PhysRevD.100.043543} {\bibfield
   {journal} {\bibinfo  {journal} {Phys. Rev. D}\ }\textbf {\bibinfo {volume}
  {100}},\ \bibinfo {pages} {043543} (\bibinfo {year} {2019})},\ \Eprint
  {https://arxiv.org/abs/1903.07487} {arXiv:1903.07487 [astro-ph.CO]}
  \BibitemShut {NoStop}%
\bibitem [{\citenamefont {Dom\`enech}(2021)}]{Domenech:2021ztg}%
  \BibitemOpen
  \bibfield  {author} {\bibinfo {author} {\bibfnamefont {G.}~\bibnamefont
  {Dom\`enech}},\ }\href {https://doi.org/10.3390/universe7110398} {\bibfield
  {journal} {\bibinfo  {journal} {Universe}\ }\textbf {\bibinfo {volume} {7}},\
  \bibinfo {pages} {398} (\bibinfo {year} {2021})},\ \Eprint
  {https://arxiv.org/abs/2109.01398} {arXiv:2109.01398 [gr-qc]} \BibitemShut
  {NoStop}%
\bibitem [{\citenamefont {Amaro-Seoane}\ \emph {et~al.}(2017)\citenamefont
  {Amaro-Seoane} \emph {et~al.}}]{LISA1}%
  \BibitemOpen
  \bibfield  {author} {\bibinfo {author} {\bibfnamefont {P.}~\bibnamefont
  {Amaro-Seoane}} \emph {et~al.}\ }\href
  {https://doi.org/10.48550/arXiv.1702.00786} {10.48550/arXiv.1702.00786}
  (\bibinfo {year} {2017})\BibitemShut {NoStop}%
\bibitem [{\citenamefont {Abbott}\ \emph {et~al.}(2016)\citenamefont {Abbott}
  \emph {et~al.}}]{LIGOScientific:2016fpe}%
  \BibitemOpen
  \bibfield  {author} {\bibinfo {author} {\bibfnamefont {B.~P.}\ \bibnamefont
  {Abbott}} \emph {et~al.} (\bibinfo {collaboration} {LIGO Scientific,
  Virgo}),\ }\href {https://doi.org/10.1103/PhysRevLett.116.131102} {\bibfield
  {journal} {\bibinfo  {journal} {Phys. Rev. Lett.}\ }\textbf {\bibinfo
  {volume} {116}},\ \bibinfo {pages} {131102} (\bibinfo {year} {2016})},\
  \Eprint {https://arxiv.org/abs/1602.03847} {arXiv:1602.03847 [gr-qc]}
  \BibitemShut {NoStop}%
\bibitem [{\citenamefont {Agazie}\ \emph {et~al.}(2023)\citenamefont {Agazie}
  \emph {et~al.}}]{NANOGrav:2023gor}%
  \BibitemOpen
  \bibfield  {author} {\bibinfo {author} {\bibfnamefont {G.}~\bibnamefont
  {Agazie}} \emph {et~al.} (\bibinfo {collaboration} {NANOGrav}),\ }\href
  {https://doi.org/10.3847/2041-8213/acdac6} {\bibfield  {journal} {\bibinfo
  {journal} {Astrophys. J. Lett.}\ }\textbf {\bibinfo {volume} {951}},\
  \bibinfo {pages} {L8} (\bibinfo {year} {2023})},\ \Eprint
  {https://arxiv.org/abs/2306.16213} {arXiv:2306.16213 [astro-ph.HE]}
  \BibitemShut {NoStop}%
\bibitem [{\citenamefont {Afzal}\ \emph {et~al.}(2023)\citenamefont {Afzal}
  \emph {et~al.}}]{NANOGrav:2023hvm}%
  \BibitemOpen
  \bibfield  {author} {\bibinfo {author} {\bibfnamefont {A.}~\bibnamefont
  {Afzal}} \emph {et~al.} (\bibinfo {collaboration} {NANOGrav}),\ }\href
  {https://doi.org/10.3847/2041-8213/acdc91} {\bibfield  {journal} {\bibinfo
  {journal} {Astrophys. J. Lett.}\ }\textbf {\bibinfo {volume} {951}},\
  \bibinfo {pages} {L11} (\bibinfo {year} {2023})},\ \Eprint
  {https://arxiv.org/abs/2306.16219} {arXiv:2306.16219 [astro-ph.HE]}
  \BibitemShut {NoStop}%
\bibitem [{\citenamefont {Antoniadis}\ \emph
  {et~al.}(2023{\natexlab{a}})\citenamefont {Antoniadis} \emph
  {et~al.}}]{EPTA:2023fyk}%
  \BibitemOpen
  \bibfield  {author} {\bibinfo {author} {\bibfnamefont {J.}~\bibnamefont
  {Antoniadis}} \emph {et~al.} (\bibinfo {collaboration} {EPTA, InPTA:}),\
  }\href {https://doi.org/10.1051/0004-6361/202346844} {\bibfield  {journal}
  {\bibinfo  {journal} {Astron. Astrophys.}\ }\textbf {\bibinfo {volume}
  {678}},\ \bibinfo {pages} {A50} (\bibinfo {year} {2023}{\natexlab{a}})},\
  \Eprint {https://arxiv.org/abs/2306.16214} {arXiv:2306.16214 [astro-ph.HE]}
  \BibitemShut {NoStop}%
\bibitem [{\citenamefont {Antoniadis}\ \emph
  {et~al.}(2023{\natexlab{b}})\citenamefont {Antoniadis} \emph
  {et~al.}}]{EPTA:2023xxk}%
  \BibitemOpen
  \bibfield  {author} {\bibinfo {author} {\bibfnamefont {J.}~\bibnamefont
  {Antoniadis}} \emph {et~al.} (\bibinfo {collaboration} {EPTA}),\ }\href@noop
  {} {\  (\bibinfo {year} {2023}{\natexlab{b}})},\ \Eprint
  {https://arxiv.org/abs/2306.16227} {arXiv:2306.16227 [astro-ph.CO]}
  \BibitemShut {NoStop}%
\bibitem [{\citenamefont {Reardon}\ \emph {et~al.}(2023)\citenamefont {Reardon}
  \emph {et~al.}}]{Reardon:2023gzh}%
  \BibitemOpen
  \bibfield  {author} {\bibinfo {author} {\bibfnamefont {D.~J.}\ \bibnamefont
  {Reardon}} \emph {et~al.},\ }\href {https://doi.org/10.3847/2041-8213/acdd02}
  {\bibfield  {journal} {\bibinfo  {journal} {Astrophys. J. Lett.}\ }\textbf
  {\bibinfo {volume} {951}},\ \bibinfo {pages} {L6} (\bibinfo {year} {2023})},\
  \Eprint {https://arxiv.org/abs/2306.16215} {arXiv:2306.16215 [astro-ph.HE]}
  \BibitemShut {NoStop}%
\bibitem [{\citenamefont {Xu}\ \emph {et~al.}(2023)\citenamefont {Xu} \emph
  {et~al.}}]{Xu:2023wog}%
  \BibitemOpen
  \bibfield  {author} {\bibinfo {author} {\bibfnamefont {H.}~\bibnamefont {Xu}}
  \emph {et~al.},\ }\href {https://doi.org/10.1088/1674-4527/acdfa5} {\bibfield
   {journal} {\bibinfo  {journal} {Res. Astron. Astrophys.}\ }\textbf {\bibinfo
  {volume} {23}},\ \bibinfo {pages} {075024} (\bibinfo {year} {2023})},\
  \Eprint {https://arxiv.org/abs/2306.16216} {arXiv:2306.16216 [astro-ph.HE]}
  \BibitemShut {NoStop}%
\bibitem [{\citenamefont {Inomata}\ and\ \citenamefont
  {Nakama}(2019)}]{Inomata:2018epa}%
  \BibitemOpen
  \bibfield  {author} {\bibinfo {author} {\bibfnamefont {K.}~\bibnamefont
  {Inomata}}\ and\ \bibinfo {author} {\bibfnamefont {T.}~\bibnamefont
  {Nakama}},\ }\href {https://doi.org/10.1103/PhysRevD.99.043511} {\bibfield
  {journal} {\bibinfo  {journal} {Phys. Rev. D}\ }\textbf {\bibinfo {volume}
  {99}},\ \bibinfo {pages} {043511} (\bibinfo {year} {2019})},\ \Eprint
  {https://arxiv.org/abs/1812.00674} {arXiv:1812.00674 [astro-ph.CO]}
  \BibitemShut {NoStop}%
\bibitem [{\citenamefont {Tytler}\ \emph {et~al.}(2000)\citenamefont {Tytler},
  \citenamefont {O'Meara}, \citenamefont {Suzuki},\ and\ \citenamefont
  {Lubin}}]{Tytler:2000qf}%
  \BibitemOpen
  \bibfield  {author} {\bibinfo {author} {\bibfnamefont {D.}~\bibnamefont
  {Tytler}}, \bibinfo {author} {\bibfnamefont {J.~M.}\ \bibnamefont {O'Meara}},
  \bibinfo {author} {\bibfnamefont {N.}~\bibnamefont {Suzuki}},\ and\ \bibinfo
  {author} {\bibfnamefont {D.}~\bibnamefont {Lubin}},\ }\href
  {https://doi.org/10.1238/Physica.Topical.085a00012} {\bibfield  {journal}
  {\bibinfo  {journal} {Phys. Scripta T}\ }\textbf {\bibinfo {volume} {85}},\
  \bibinfo {pages} {12} (\bibinfo {year} {2000})},\ \Eprint
  {https://arxiv.org/abs/astro-ph/0001318} {arXiv:astro-ph/0001318}
  \BibitemShut {NoStop}%
\end{thebibliography}%

\clearpage
\newpage
\maketitle
\onecolumngrid
\begin{center}
\textbf{\large Inflationary Butterfly Effect: \\ Non-Perturbative Dynamics From Small-Scale Features } \\ 
\vspace{0.05in}
{ \it \large Supplemental Material}\\ 
\vspace{0.05in}
{Angelo Caravano, Keisuke Inomata, and S\'ebastien Renaux-Petel}
\end{center}
\onecolumngrid
\setcounter{equation}{0}
\setcounter{figure}{0}
\setcounter{table}{0}
\setcounter{section}{0}
\setcounter{page}{1}
\makeatletter
\renewcommand{\theequation}{S\arabic{equation}}
\renewcommand{\thefigure}{S\arabic{figure}}

\setcounter{secnumdepth}{1}

\section{Fraction of primordial black holes produced during an early matter era}
\label{smsec:pbh}

Here we explain why the PBH mass fraction $\beta$ can be as large as $\beta \sim 10^{-6}$ without PBH overclosing the Universe, if we consider PBHs produced during a MD era. 
If the Universe experiences an early MD (eMD) era before big bang nucleosynthesis (BBN) and during which PBHs are produced, the fraction of PBHs to the total energy density is constant until the end of the eMD era. 
We consider here that the eMD era ends with entropy production through the decay of the dominant component to radiation, and is then followed by the RD era where BBN occurs. 
Once the RD era begins, the fraction of PBHs to the total energy density increases proportionally to the scale factor. 
This is because the energy density of PBHs is $\propto a^{-3}$, while the total energy density is $\propto a^{-4}$ during the RD era. 
Thus, the current fraction of PBHs to dark matter, $f_\pbh$, is related to the mass fraction of PBHs
at formation, $\beta$, as 
\begin{align}
    f_\pbh \simeq \frac{\Omega_\text{DM}}{\Omega_\text{DM} + \Omega_\text{b}}\frac{a_\eq}{a_*} \beta \simeq \frac{\Omega_\text{DM}}{\Omega_\text{DM} + \Omega_\text{b}}\frac{T_*}{T_\eq} \beta,
    \label{eq:fpbh}
\end{align}
where $\Omega_\text{DM}$ and $\Omega_\text{b}$ are the energy density parameters for dark matter and baryon, and the subscript $*$ and $\eq$ denote, respectively, the values at the end of the eMD era, and at the time of the matter-radiation equality ($T_\eq \simeq 8.0\times 10^{-7} \,\text{MeV}$~\cite{Inomata:2018epa}).
Note that for simplicity, we neglect the change in the number of degrees of freedom, which does 
not affect the order of magnitude of the resulting upper bound on $\beta$.

The eMD era must end before BBN, which occurs around $T \simeq \mathcal O(1)\, \text{MeV}$~\cite{Tytler:2000qf}. Otherwise, the predictions of BBN would be modified and inconsistent with the observation of light elements in the Universe. 
From this, we deduce the lower bound $T_* >\mathcal O(1)\, \text{MeV}$.
Substituting the concrete values of $T_\eq$ and $T_*$ into Eq.~(\ref{eq:fpbh}), we obtain 
$f_\pbh \gtrsim \mathcal O(10^6) \beta$. Imposing $f_\pbh < 1$, we finally obtain the upper bound on $\beta$:
\begin{align}
    \beta \lesssim \mathcal O(10^{-6}).
\end{align}

\end{document}